\documentclass[useAMS,usenatbib]{mn2e}
\usepackage{graphicx,natbib}

\def\aV{\mbox{$\rm A_V$}}

\def\jh{\mbox{$(J-H)$}}
\def\hk{\mbox{$(H-K_s)$}}
\def\jk{\mbox{$(J-K_s)$}}

\def\ebv{\mbox{$E(B-V)$}}

\def\ejh{\mbox{$E(J-H)$}}

\def\rc{\mbox{$R_{\rm c}$}}

\def\rl{\mbox{$R_{\rm RDP}$}}
\def\rx{\mbox{$R_{\rm ext}$}}

\def\ds{\mbox{$d_\odot$}}

\def\zgc{\mbox{$Z_{\rm GP}$}}
\def\dSC{\mbox{$\Delta R_{\rm SC}$}}

\def\jj{\mbox{$J$}}
\def\hh{\mbox{$H$}}
\def\ks{\mbox{$K_s$}}

\title[Overlooked Ruprecht clusters]{Characterisation of 15 overlooked Ruprecht 
clusters with ages within 400\,Myr and 3\,Gyr.}

\author[C. Bonatto and E. Bica]{C. Bonatto$^1$ and E. Bica$^1$\\
$^1$ Departamento de Astronomia, Universidade Federal do Rio Grande do Sul, 
Av. Bento Gon\c{c}alves 9500\\
Porto Alegre 91501-970, RS, Brazil}

\begin{document}

\pagerange{\pageref{firstpage}--\pageref{lastpage}}

\maketitle

\label{firstpage}

\begin{abstract}
We derive fundamental, structural, and photometric parameters of 15 overlooked 
Ruprecht (hereafter Ru) star clusters by means of 2MASS photometry and field-star 
decontamination. Ru\,1, 10, 23, 26, 27, 34, 35, 37, 41, 54, 60, 63, 66, and 152 
are located in the third Galactic quadrant, while Ru\,174 is in the first. With 
the constraints imposed by the field-decontaminated colour-magnitude diagrams 
(CMDs) and stellar radial density profiles (RDPs), we derive ages in the range 
400\,Myr --- 1\,Gyr, except for the older Ru\,37, with $\sim3$\,Gyr. Distances 
from the Sun are within $\rm1.5\la\ds(kpc)\la8.0$. The RDPs are well-defined and 
can be described by a King-like profile for most of the radial range, except for 
Ru\,23, 27, 41, 63, and 174, which present a conspicuous stellar density excess 
in the central region. The clusters dwell between (or close to) the Perseus and
Sagittarius-Carina arms. We derive evidence in favour of cluster size increasing
with distance to the Galactic plane ($\zgc$), which is consistent with a low 
frequency of tidal stress associated with high-$|\zgc|$ regions. The clusters are 
rather faint even in the near-infrared, with apparent integrated \jj\ magnitudes 
within $6.4\la m_J\la9.8$, while their absolute magnitudes are $-6.6\la M_J\la-2.6$. 
Extrapolation of the relation between $M_V$ and $M_J$, derived for globular clusters,
suggests that they are low-luminosity optical clusters, with $-5\la M_V\la-1$.
\end{abstract}


\section{Introduction}
\label{Intro}

In general, open clusters (OCs) are formed and evolve in - or close to - the 
Galactic disk. As a consequence of their orbits, they are constantly suffering
tidal stress from Galactic substructures and undergoing different degrees of mass 
loss that, in most cases, might lead to cluster dissolution. Indeed, the vast 
majority of the OCs do not survive even the embedded phase (e.g. \citealt{LL2003}), 
and few reach ages older than $\sim10^8$\,yr (e.g. \citealt{GoBa06}).

As systems in which the balance between velocity dispersion and escape
velocity plays a vital role, OC stellar distributions change continually
as a function of time because of mass segregation and evaporation, tidal 
interactions with the disk and/or bulge, encounters with giant molecular 
clouds, as well as mass loss associated with stellar evolution. On average, 
these processes tend to accelerate the 
cluster dynamical evolution and change the internal structure in varying 
degrees. Indeed, near the solar circle, theoretical and observational evidence 
(e.g. \citealt{Spitzer58}; \citealt{Oort58}; \citealt{BM03}; \citealt{GoBa06};
\citealt{LG06}; \citealt{Khalisi07}; \citealt{Piskunov07}) suggest a mass 
dependent disruption time scale of a few $10^8$\,yr. As a consequence, most OCs 
end up completely dissolved in the Galactic stellar field (e.g. \citealt{Lamers05}) 
or leave only poorly-populated remnants (\citealt{PB07} and references therein),
long before reaching $\sim1$\,Gyr of age.

Reflecting the above age/dissolution effect, only $\approx13\%$ of the $\approx1100$ 
OCs with known age listed in WEBDA\footnote{\em obswww.univie.ac.at/webda} are older 
than 1\,Gyr, while $\approx45\%$ have an age between 100\,Myr and 1\,Gyr, and 
$\approx42\%$ are younger than 100\,Myr. So, besides the obvious importance of 
deriving reliable astrophysical parameters for as yet unstudied clusters, the 
unambiguous characterisation of OCs older than several $10^8$\,yr will increase the
statistics of objects undergoing the dissolution phase. This, in turn, can be used 
for better determining the time scale for cluster dissolution in the Galaxy.

Cluster databases such as WEBDA still contain many unstudied objects that, over
the years, have been identified as star cluster candidates, usually based on the 
appearance in optical images. Not surprisingly, close investigations of some of 
these overlooked objects have uncovered a number of actual OCs (e.g. \citealt{Carr05}; 
\citealt{Carr06a}; \citealt{Carr06b}). Besides, since young OCs are relatively 
easy to identify (because of the presence of bright stars) even at relatively large
distances, the overlooked clusters, in general, are as old as several $10^8$\,yr, 
sometimes reaching a few Gyr (e.g. the $\sim4$\,Gyr old OC Berkeley\,56 - \citealt{Carr06b}).

\begin{figure*}
\begin{minipage}[b]{0.33\linewidth}
\includegraphics[width=\textwidth]{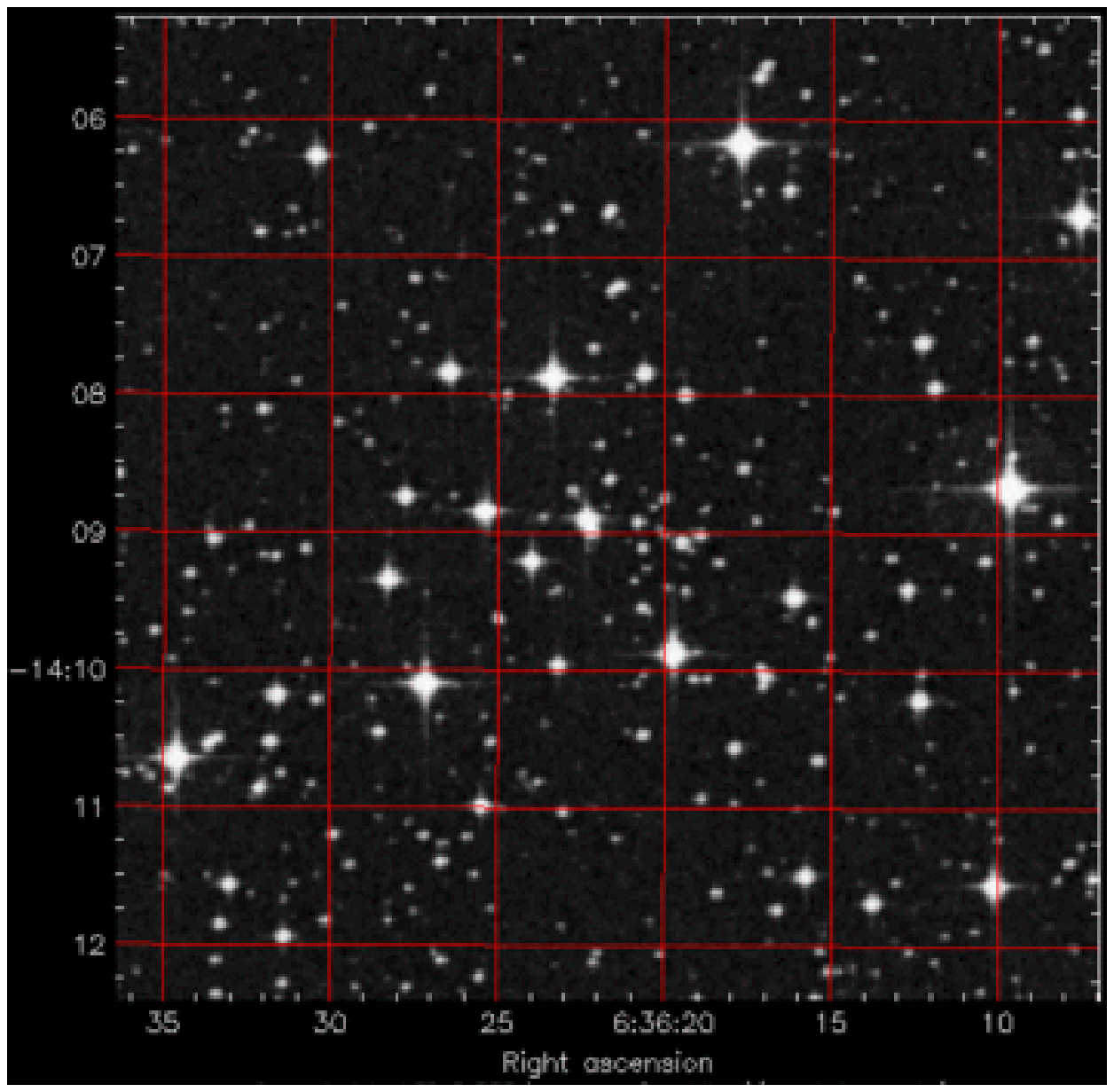}
\end{minipage}\hfill
\begin{minipage}[b]{0.33\linewidth}
\includegraphics[width=\textwidth]{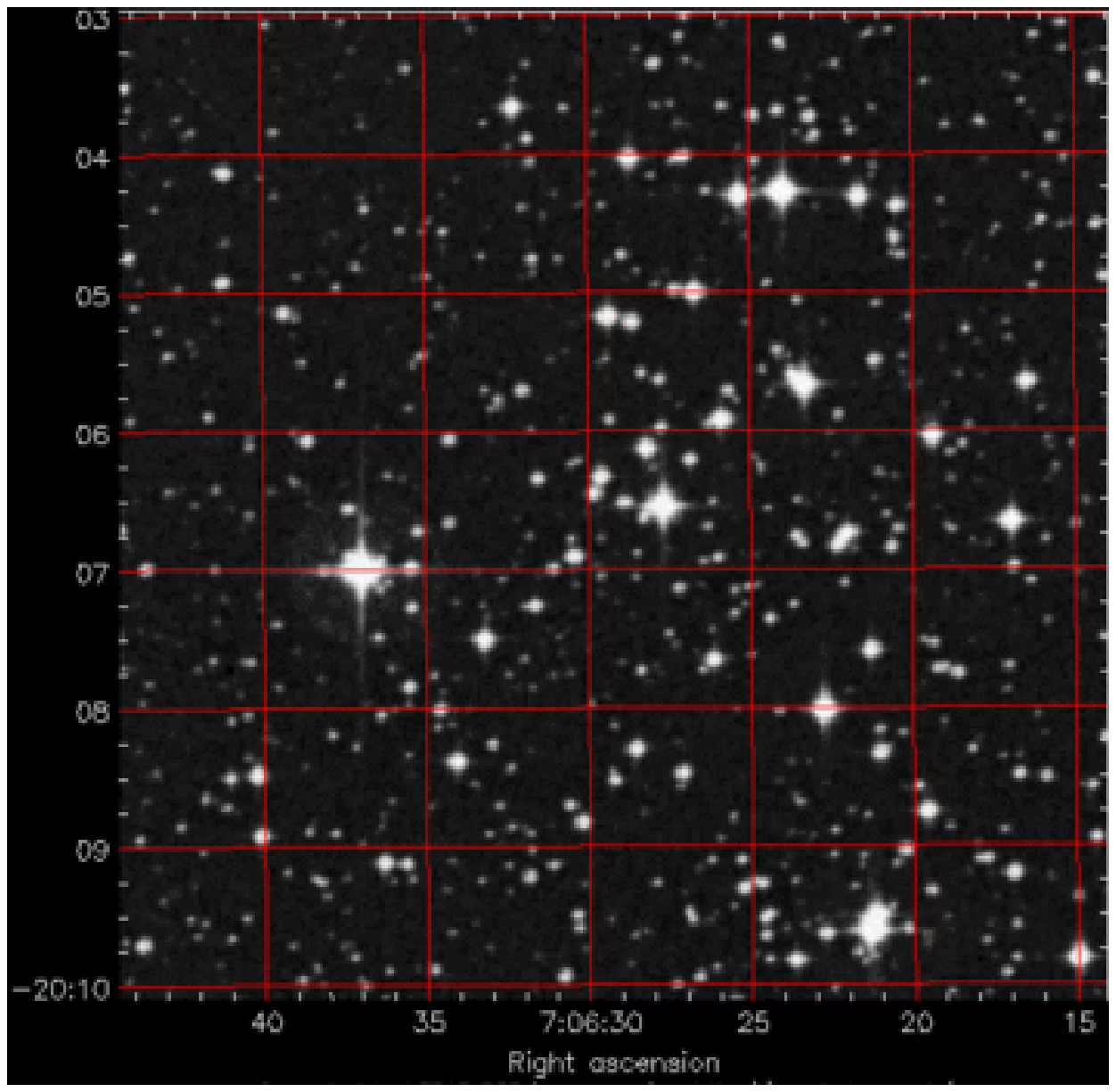}
\end{minipage}\hfill
\begin{minipage}[b]{0.33\linewidth}
\includegraphics[width=\textwidth]{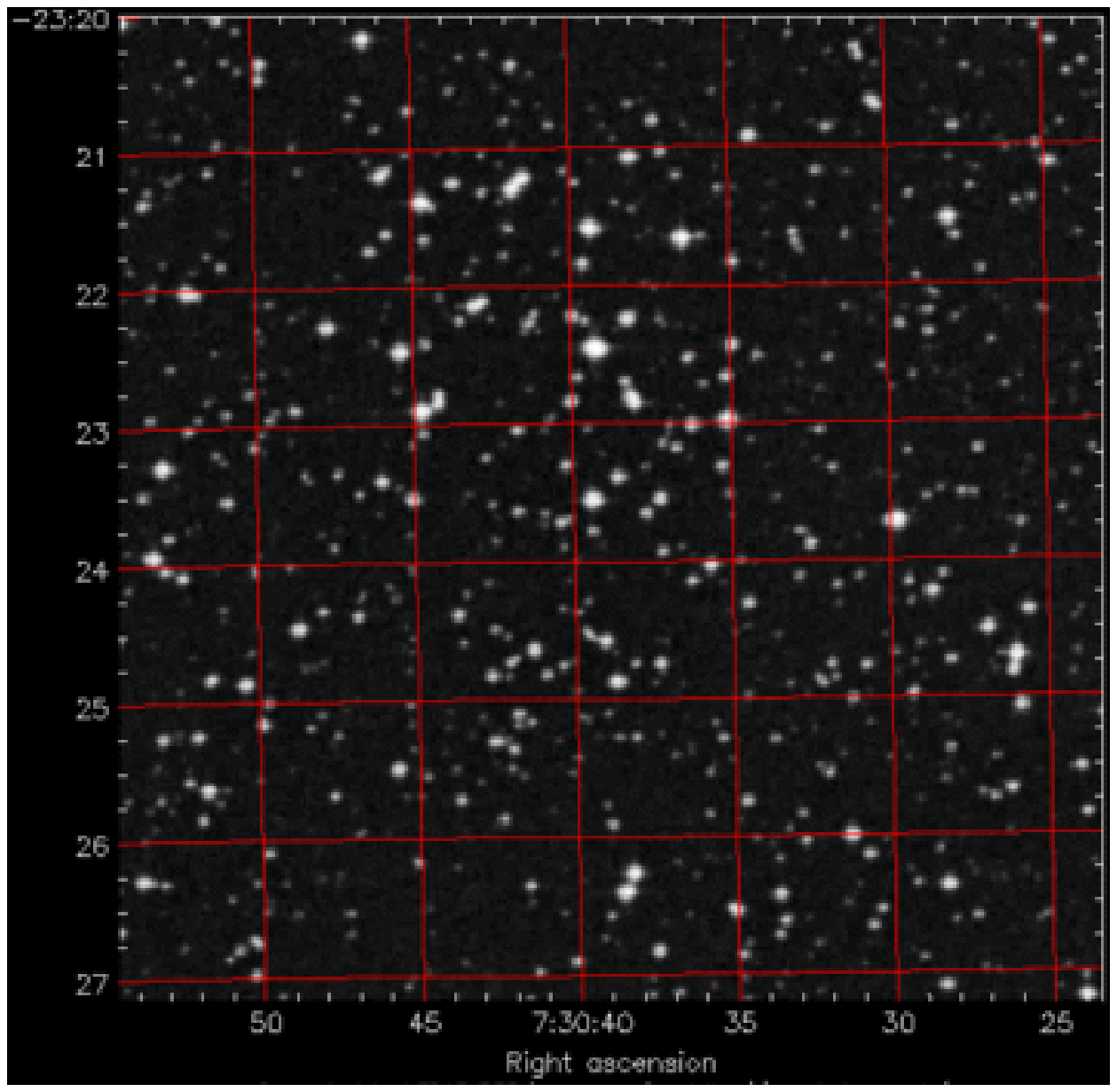}
\end{minipage}\hfill
\begin{minipage}[b]{0.33\linewidth}
\includegraphics[width=\textwidth]{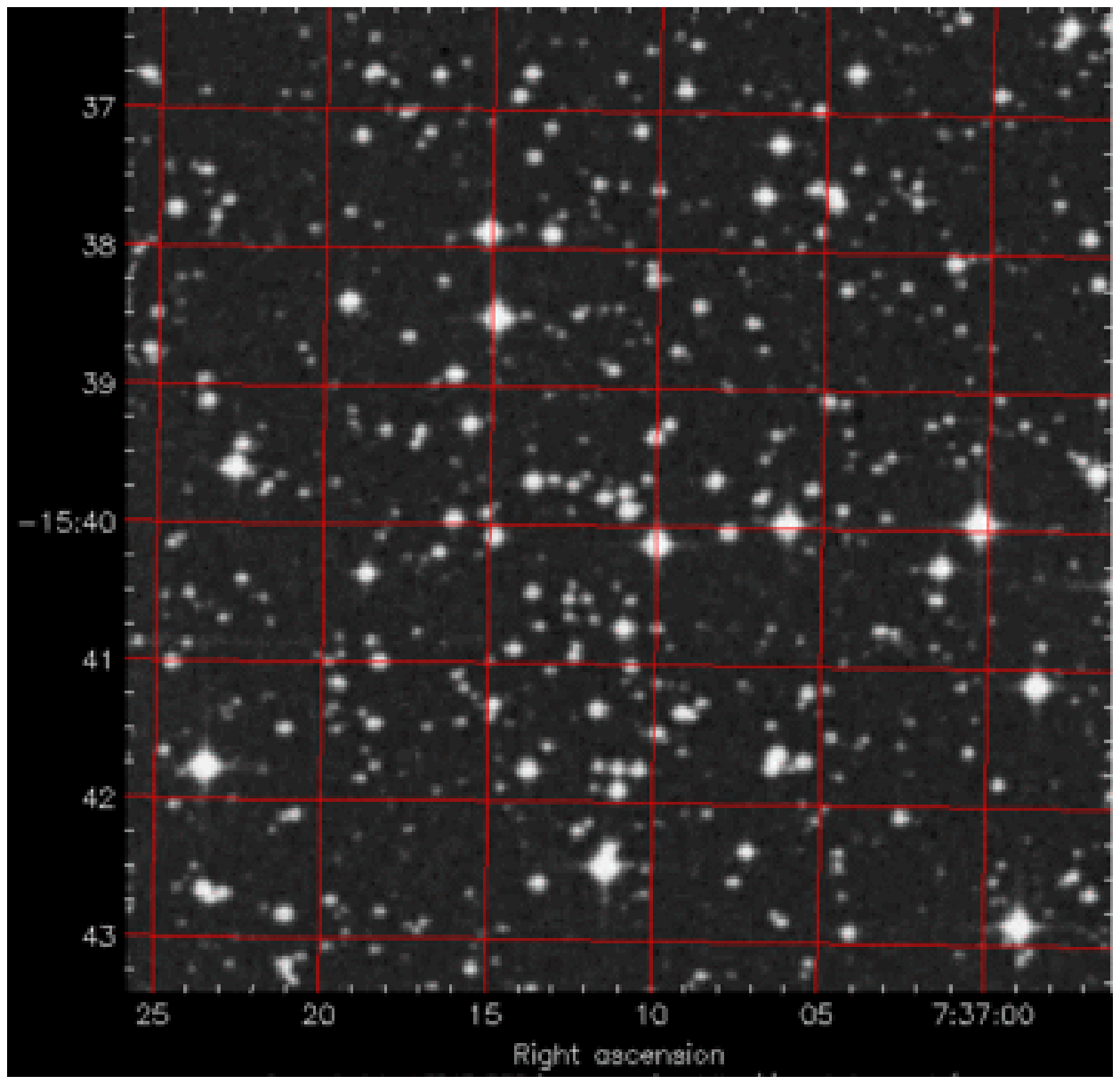}
\end{minipage}\hfill
\begin{minipage}[b]{0.33\linewidth}
\includegraphics[width=\textwidth]{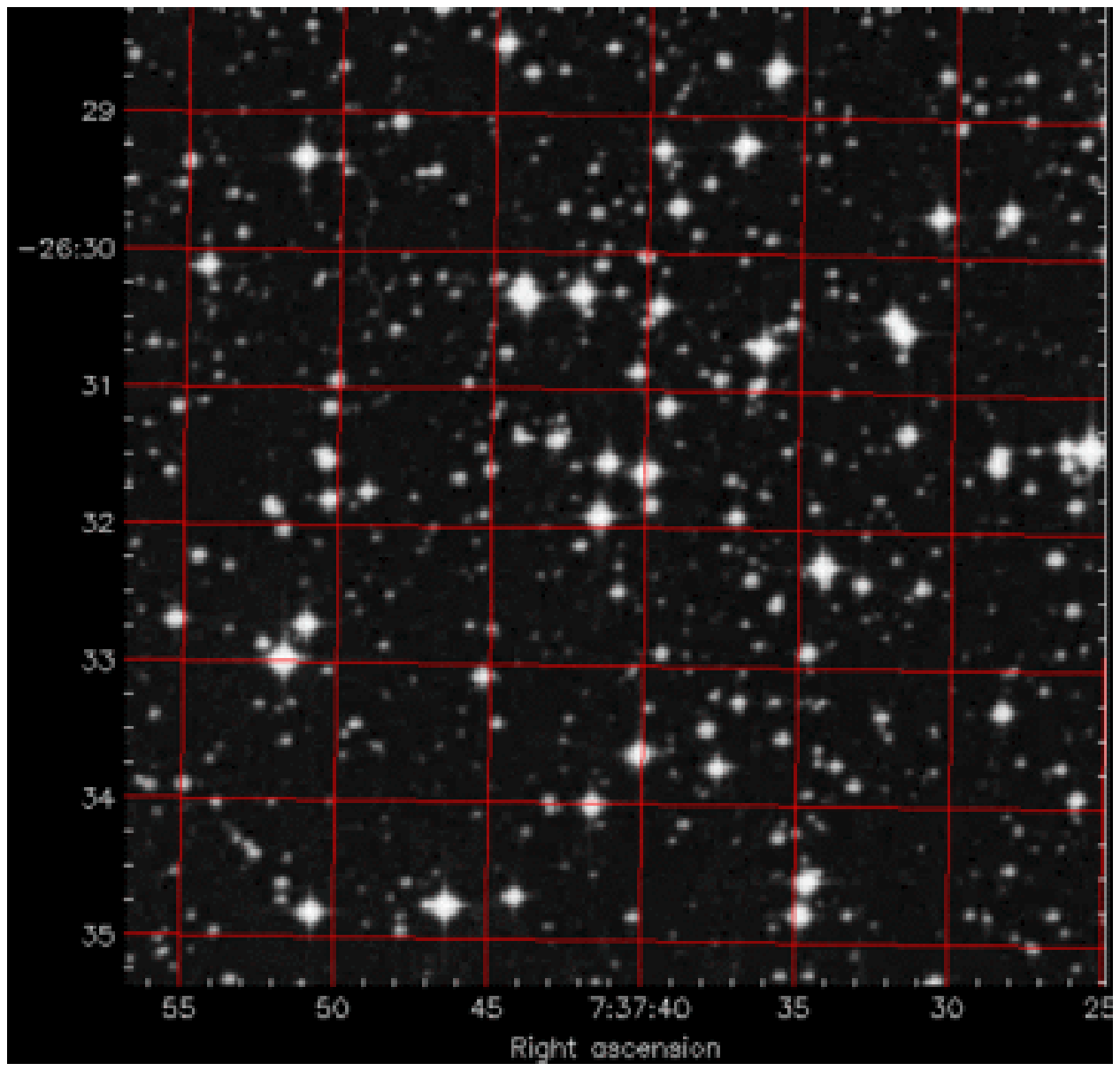}
\end{minipage}\hfill
\begin{minipage}[b]{0.33\linewidth}
\includegraphics[width=\textwidth]{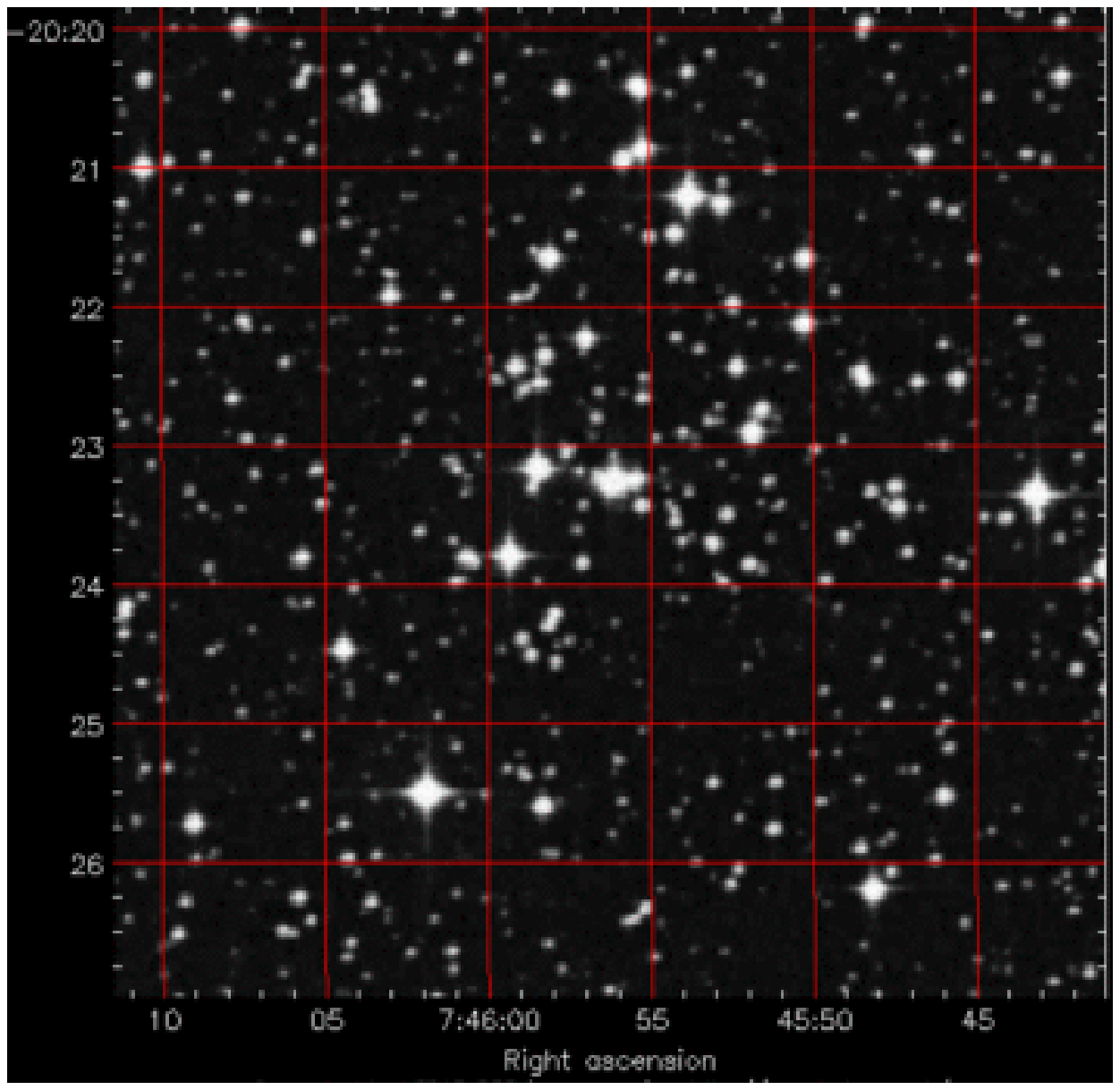}
\end{minipage}\hfill
\begin{minipage}[b]{0.33\linewidth}
\includegraphics[width=\textwidth]{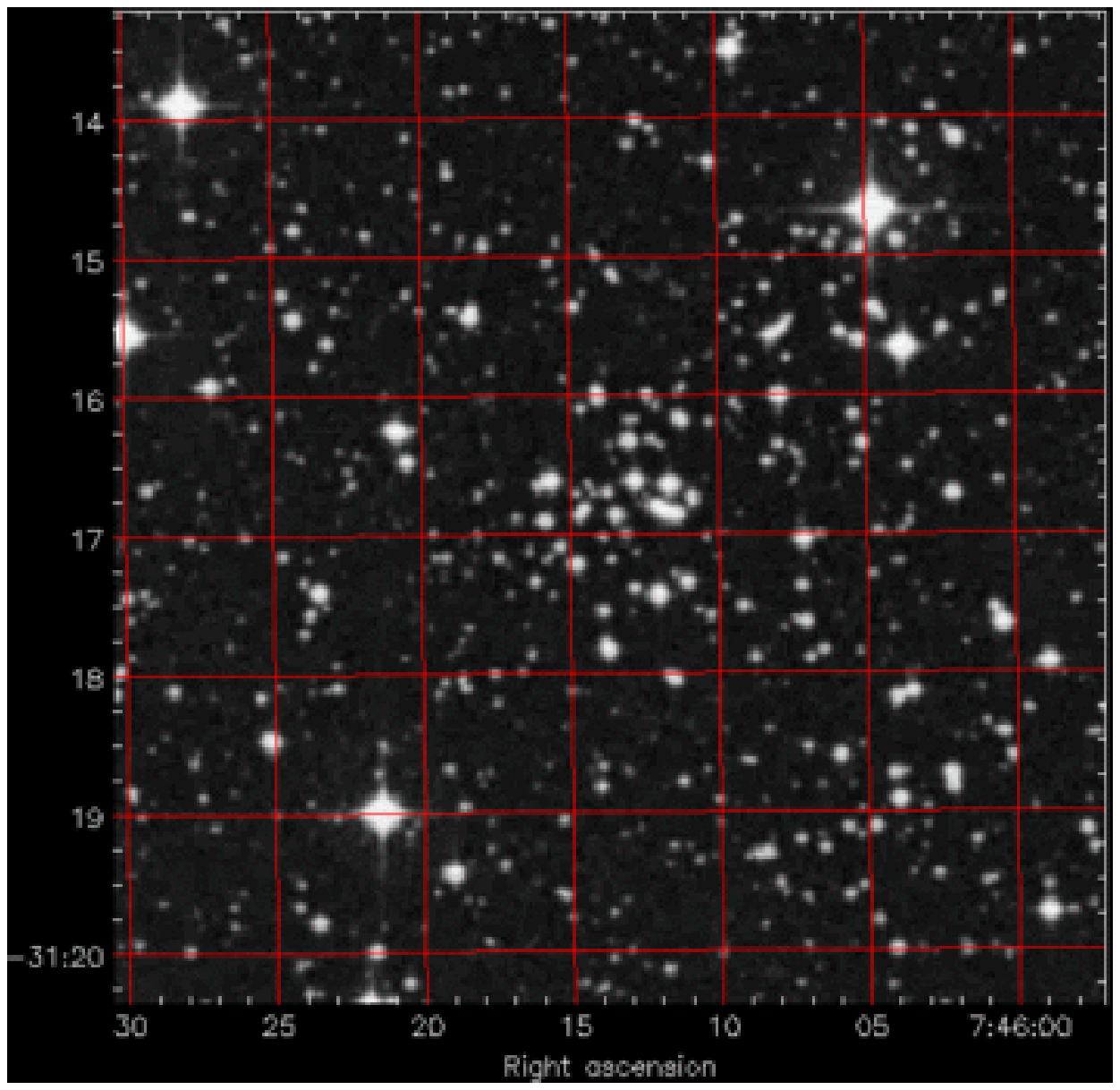}
\end{minipage}\hfill
\begin{minipage}[b]{0.33\linewidth}
\includegraphics[width=\textwidth]{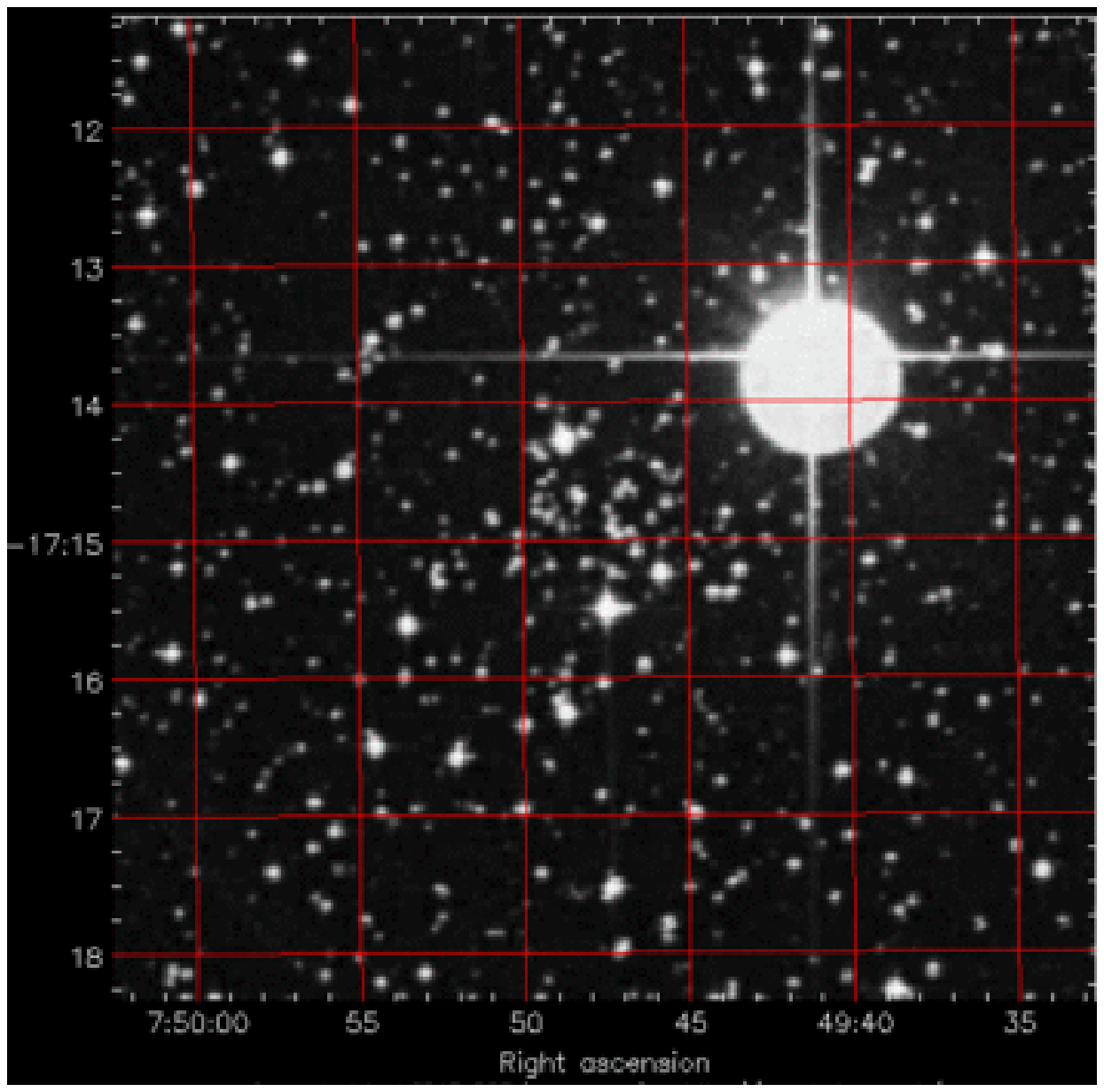}
\end{minipage}\hfill
\begin{minipage}[b]{0.33\linewidth}
\includegraphics[width=\textwidth]{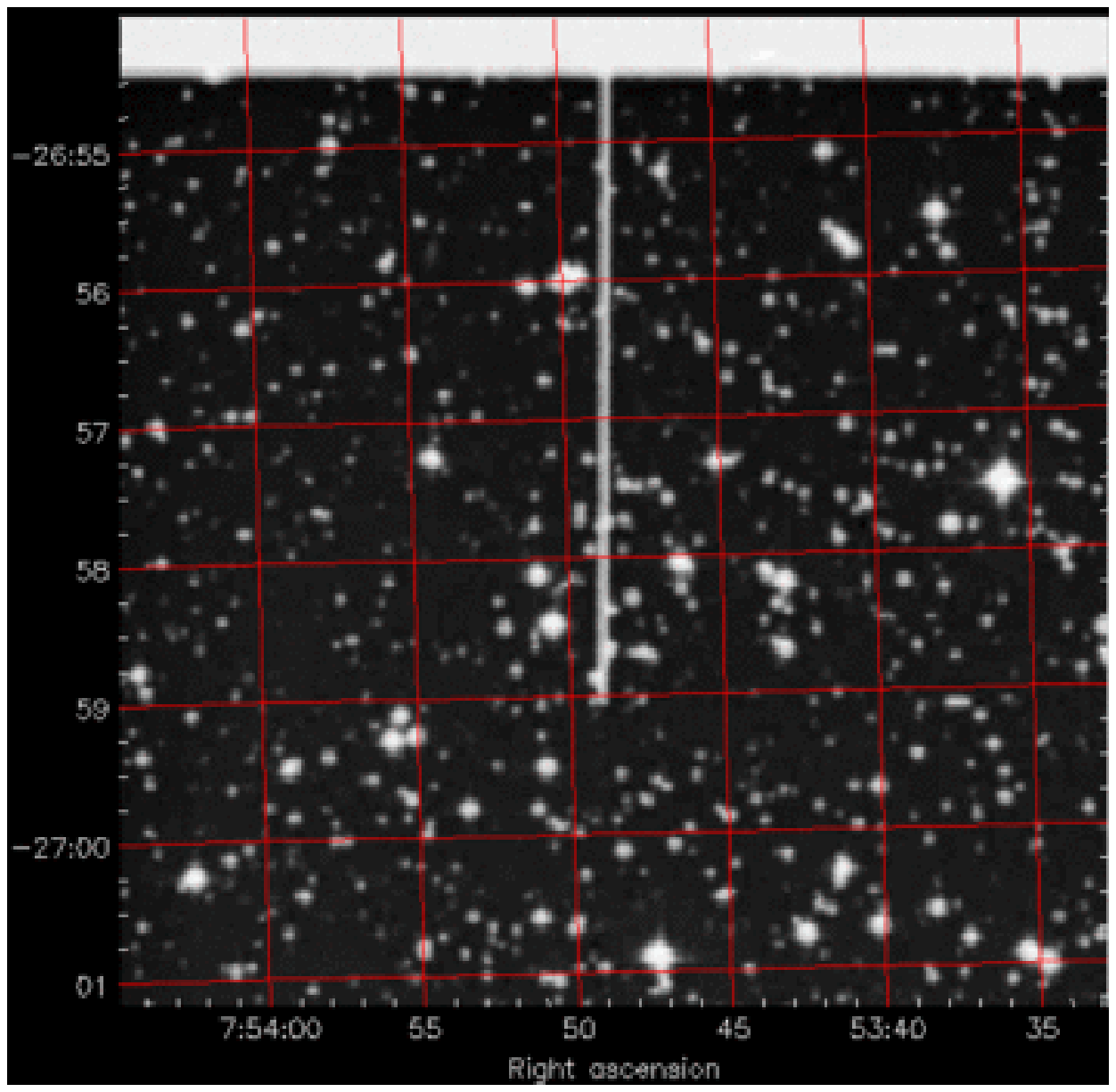}
\end{minipage}\hfill
\caption[]{All panels display $5\arcmin\times5\arcmin$ DSS-I B images. From
left to right: Ru\,1, 10, and 23 (top); Ru\,26, 27, and 34 
(middle); Ru\,35, 37 (with the bright K\,3\,III star 6\,Pup at $\approx2\arcmin$
to the northwest), and Ru\,41 (bottom). Plate flaws show up in the image of Ru\,41.
Orientation: North at the top and east at left.}
\label{fig1}
\end{figure*}

\begin{figure*}
\begin{minipage}[b]{0.33\linewidth}
\includegraphics[width=\textwidth]{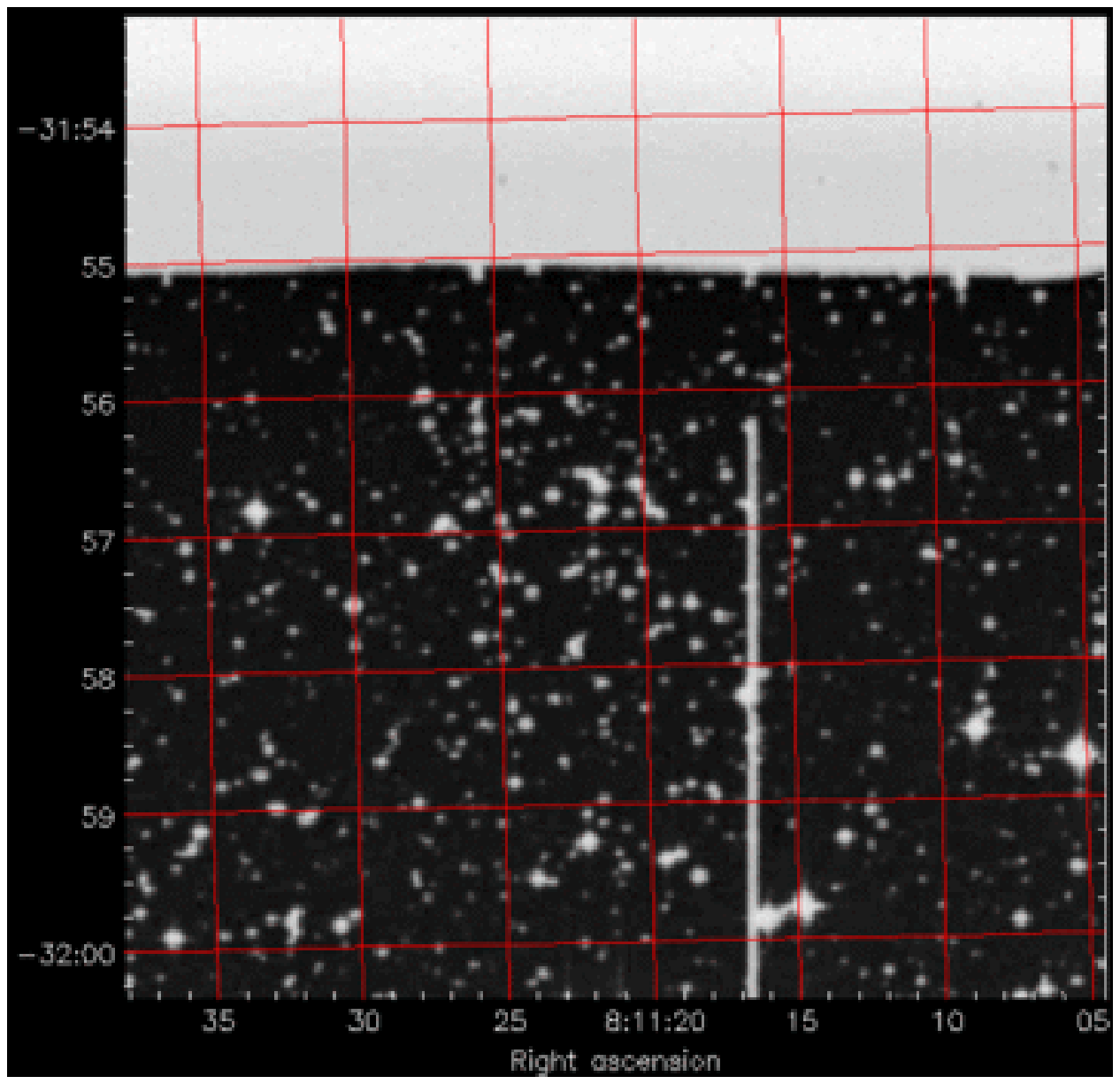}
\end{minipage}\hfill
\begin{minipage}[b]{0.33\linewidth}
\includegraphics[width=\textwidth]{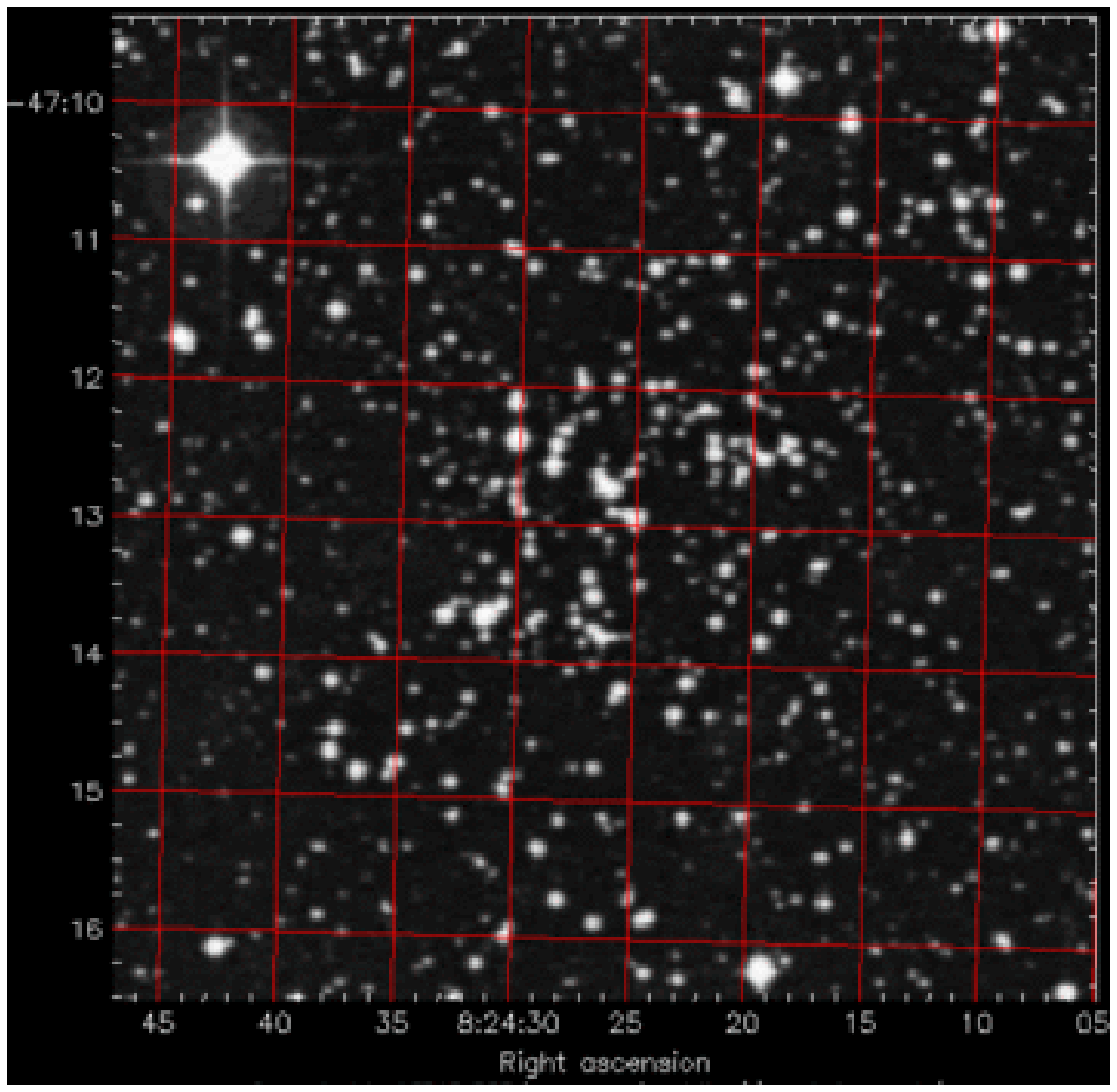}
\end{minipage}\hfill
\begin{minipage}[b]{0.33\linewidth}
\includegraphics[width=\textwidth]{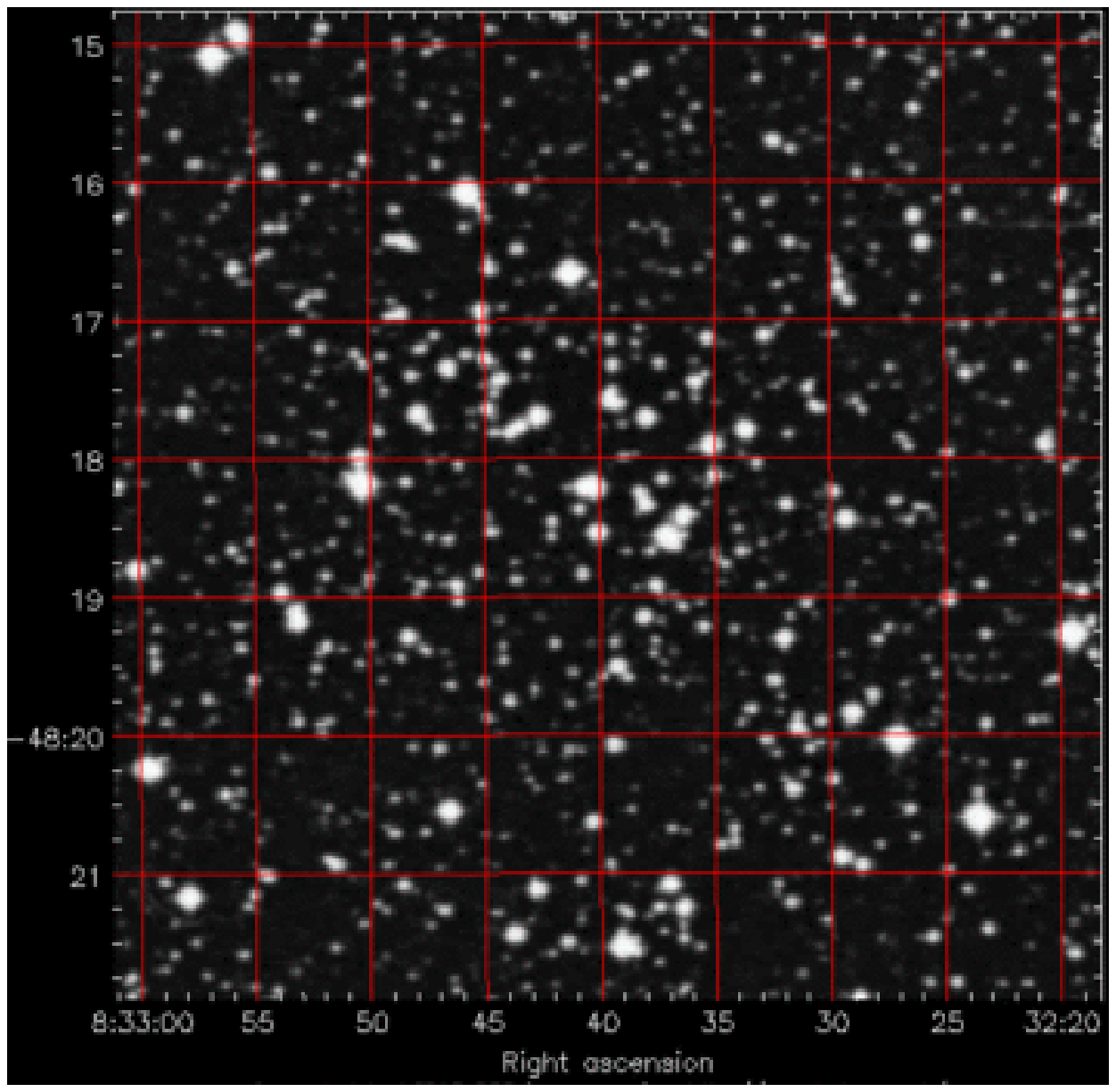}
\end{minipage}\hfill
\begin{minipage}[b]{0.33\linewidth}
\includegraphics[width=\textwidth]{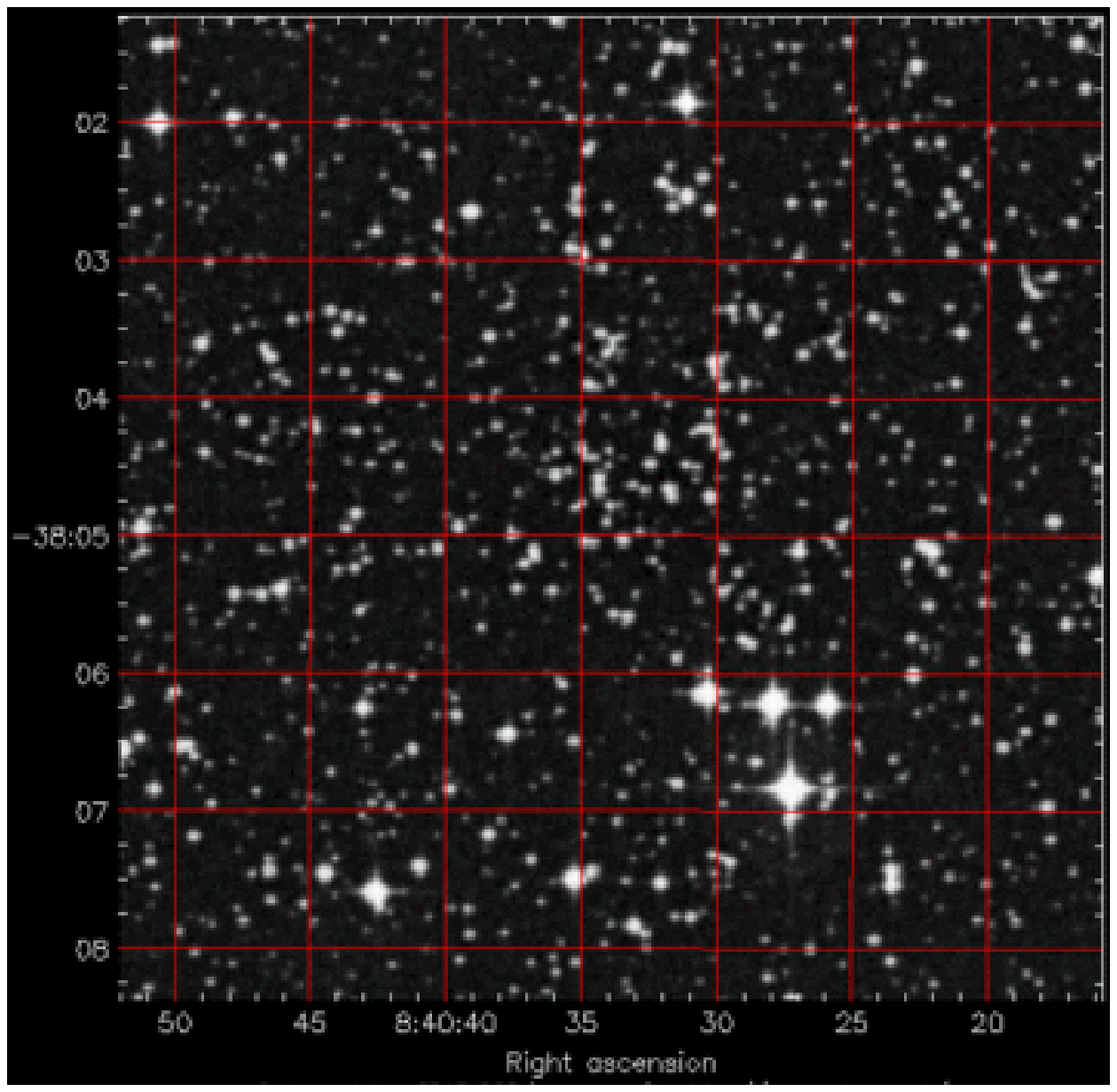}
\end{minipage}\hfill
\begin{minipage}[b]{0.33\linewidth}
\includegraphics[width=\textwidth]{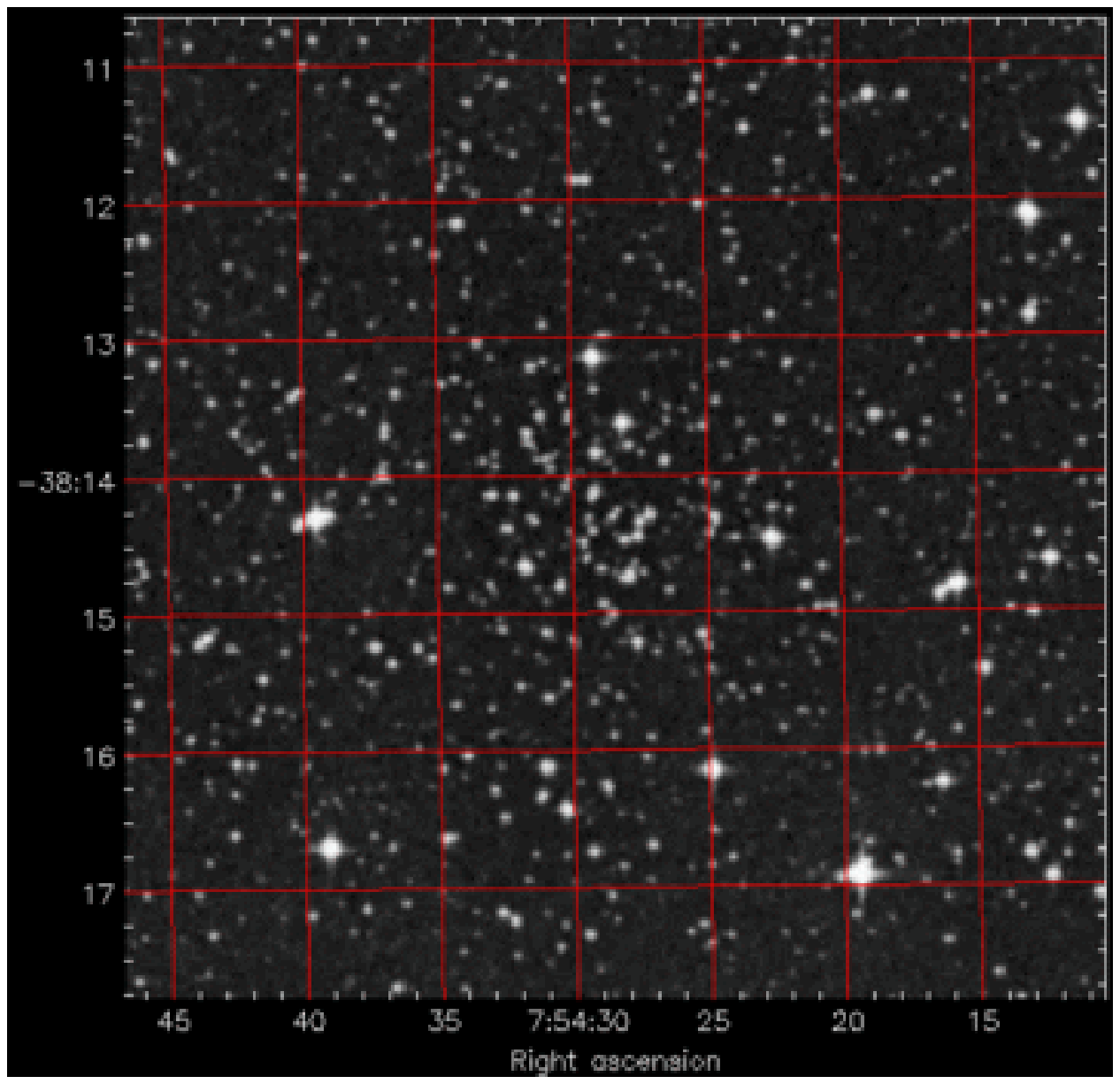}
\end{minipage}\hfill
\begin{minipage}[b]{0.33\linewidth}
\includegraphics[width=\textwidth]{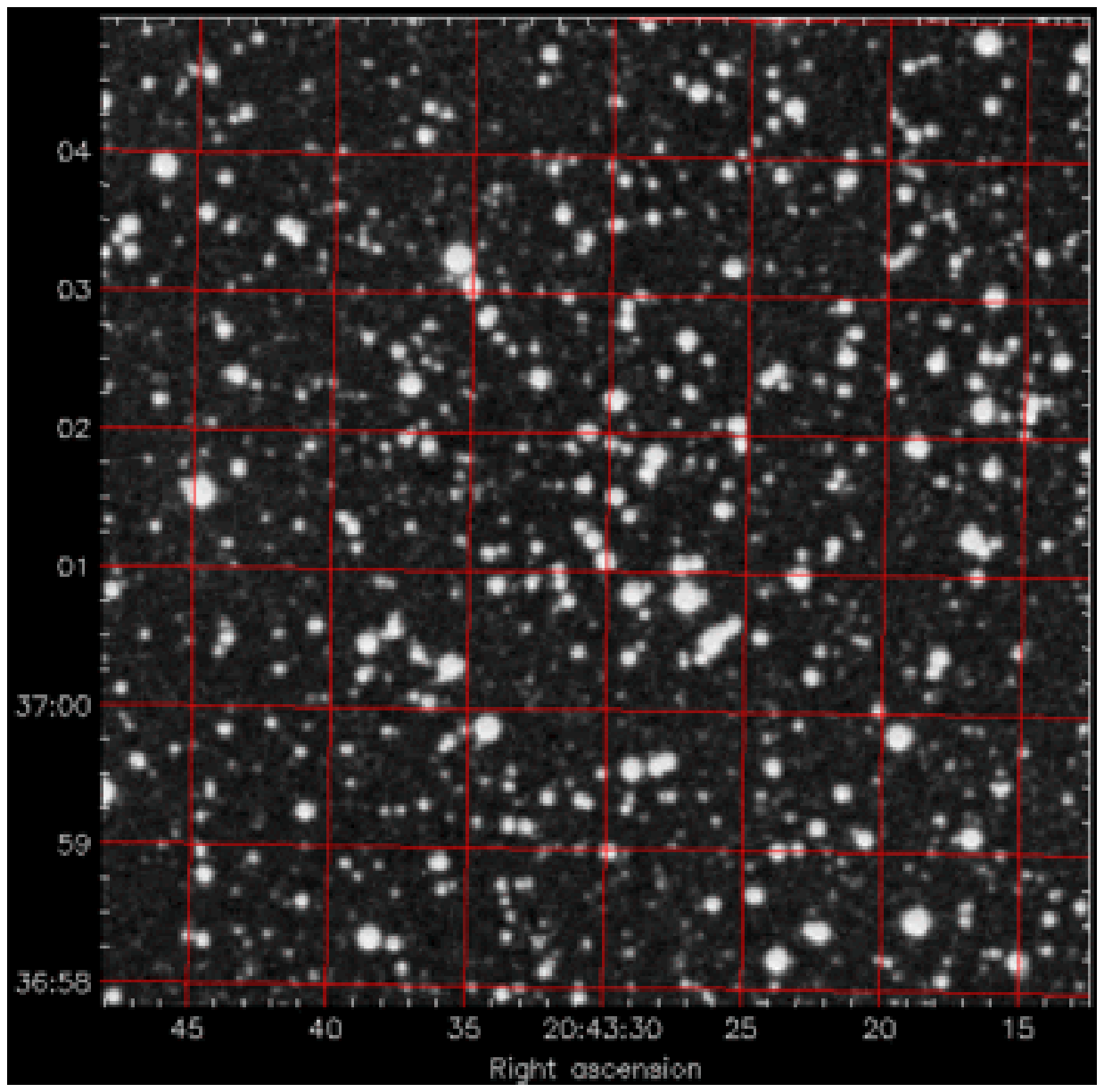}
\end{minipage}\hfill
\caption[]{Same as Fig.~\ref{fig1} for the remaining clusters. From left to 
right: Ru\,54, 60, and 63 (top); Ru\,66, 152, and 174 (bottom). Plate flaws 
show up in the image of Ru\,54.}
\label{fig2}
\end{figure*}

Based on the optical {\em Catalogue of Star Clusters and Associations} of 
\citet{Alter70}, WEBDA lists the coordinates of 171 OC candidates originally
found by J. Ruprecht (classified as Ruprecht clusters; hereafter Ru). However, 
only 79 have been so far confirmed as OCs, having 
determinations of fundamental parameters, such as the age, reddening, and 
distance from the Sun. In the present paper we derive fundamental and structural 
parameters for 15 of these overlooked Ruprecht OCs, 12 of these not 
previously studied, and the remaining 3 with inconsistent determinations 
(Sects.~\ref{DFP} and \ref{struc}). Steps taken in the present work can be summarised 
as follows: {\em (i)} 2MASS\footnote{The Two Micron All Sky Survey, All Sky data 
release (\citealt{2mass1997}) - 
{\em http://www.ipac.caltech.edu/2mass/releases/allsky/}} photometry is extracted
in a wide circular region centred on a given cluster, {\em (ii)} we apply 
field-star decontamination to uncover the intrinsic colour-magnitude diagram
(CMD) morphology, which is essential for a proper derivation of reddening, age, 
and distance from the Sun, and {\em (iii)} we apply colour-magnitude filters
to produce more contrasted  stellar radial density profiles (RDPs). In particular, 
field-star decontaminated CMDs constrain the fundamental parameters more than the 
raw (observed) photometry, especially for low-latitude and/or bulge-projected 
OCs (e.g. \citealt{ProbFSR}, and references therein).

\begin{table*}
\caption[]{Fundamental parameters}
\label{tab1}
\tiny
\renewcommand{\tabcolsep}{1.1mm}
\renewcommand{\arraystretch}{1.25}
\begin{tabular}{cccccccccccccccc}
\hline\hline
&\multicolumn{6}{c}{WEBDA}&&\multicolumn{8}{c}{This work}\\
\cline{2-7}\cline{9-16}
Cluster&$\alpha(2000)$&$\delta(2000)$&Age&\ebv&\ds&D&&$\alpha(2000)$&$\delta(2000)$&
        $\ell$&$b$&Age&\ebv&\ds&\dSC\\
 & (hms)&($\degr\,\arcmin\,\arcsec$)&(Myr)&(mag)&(kpc)&(\arcmin)&&(hms)&($\degr\,\arcmin\,\arcsec$)
 &(\degr)&(\degr)&(Myr)&(mag)&(kpc)&(kpc)\\
(1)&(2)&(3)&(4)&(5)&(6)&(7)&&(8)&(9)&(10)&(11)&(12)&(13)&(14)&(15)\\
\hline
Ru\,1  &06:36:25&$-$14:10:48&580&0.15&1.1&5.0&&06:36:21.63&$-$14:08:48.75&223.95&$-$9.68&$500\pm100$&$0.26\pm0.06$&$1.72\pm0.16$&$1.31\pm0.12$\\
Ru\,10 &07:06:25&$-$20:05:00&---&---&--- &4.0&&07:06:29.05&$-$20:06:31.50&232.55&$-$5.85&$500\pm100$&$0.64\pm0.06$&$2.34\pm0.22$&$1.62\pm0.14$ \\
Ru\,23 &07:30:41&$-$23:23:00&---&---&--- &4.0&&07:30:38.80&$-$23:23:36.00&238.08&$-$2.40&$600\pm100$&$0.54\pm0.06$&$3.06\pm0.29$&$2.00\pm0.16$ \\
Ru\,26 &07:37:11&$-$15:38:59&30&0.10&1.4&24.0&&07:37:11.10&$-$15:39:46.50&232.06&$+$2.68&$400\pm50$ &$0.35\pm0.06$&$1.82\pm0.17$&$1.24\pm0.11$ \\
Ru\,27 &07:37:30&$-$26:31:42&250&0.15&0.6&21.6&&07:37:40.89&$-$26:31:45.50&241.60&$-$2.53&$900\pm100$&$0.03\pm0.06$&$1.49\pm0.14$&$0.82\pm0.07$ \\
Ru\,34 &07:45:55&$-$20:23:00&---&---&--- &6.0&&07:45:56.22&$-$20:23:24.00&237.20&$+$2.16&$1000\pm100$&$0.00\pm0.06$&$2.63\pm0.25$&$1.70\pm0.14$ \\
Ru\,35 &07:46:13&$-$31:17:00&---&---&--- &1.5&&07:46:13.64&$-$31:16:47.25&246.66&$-$3.25&$400\pm100$&$0.45\pm0.10$&$3.91\pm0.56$&$2.26\pm0.28$ \\
Ru\,37 &07:49:54&$-$17:17:00&---&---&--- &4.0&&07:49:47.53&$-$17:14:46.50&234.94&$+$4.53&$3000\pm1000$&$0.00\pm0.06$&$5.25\pm0.74$&$3.87\pm0.45$ \\
Ru\,41 &07:53:49&$-$26:58:00&---&---&--- &2.0&&07:53:48.38&$-$26:57:39.00&243.78&$+$0.37&$700\pm100$&$0.13\pm0.10$&$3.15\pm0.45$&$1.84\pm0.23$ \\
Ru\,54 &08:11:21&$-$31:57:00&---&---&--- &3.0&&08:11:20.92&$-$31:56:49.50&250.03&$+$0.96&$800\pm100$&$0.13\pm0.10$&$5.47\pm0.78$&$3.22\pm0.43$ \\
Ru\,60 &08:24:27&$-$47:13:00&---&---&--- &3.0&&08:24:26.29&$-$47:12:51.01&264.10&$-$5.51&$400\pm100$&$0.64\pm0.10$&$6.16\pm0.88$&$2.74\pm0.54$ \\
Ru\,63 &08:32:40&$-$48:18:00&---&---&--- &3.0&&08:32:39.60&$-$48:18:19.50&265.80&$-$5.02&$500\pm100$&$0.61\pm0.10$&$3.76\pm0.37$&$1.16\pm0.17$ \\
Ru\,66 &08:40:33&$-$38:04:00&---&---&--- &2.0&&08:40:33.82&$-$38:04:47.99&258.49&$+$2.28&$600\pm100$&$0.90\pm0.10$&$3.76\pm0.54$&$1.56\pm0.24$ \\
Ru\,152&07:54:30&$-$38:14:00&---&---&--- &3.0&&07:54:28.35&$-$38:14:14.26&253.54&$-$5.30&$600\pm100$&$0.67\pm0.10$&$8.02\pm1.15$&$5.00\pm0.73$ \\
Ru\,174&20:43:30&$+$37:03:00&---&---&--- &8.0&&20:43:30.48&$+$37:01:26.26& 78.02&$-$3.39&$800\pm100$&$0.32\pm0.06$&$2.11\pm0.20$&$-0.13\pm0.07$ \\
\hline
\end{tabular}
\begin{list}{Table Notes.}
\item Col.~7: diameter estimated from optical images (\citealt{diasCat}); Col.~14: distance 
from the Sun; col.~15: distance from the Solar circle. 
\end{list}
\end{table*}

This paper is organised as follows. In Sect.~\ref{RecAdd} we present optical images
of the sample objects. In Sect.~\ref{2mass} we discuss the 2MASS photometry and build 
the field-star decontaminated CMDs. In Sect.~\ref{DFP} we derive fundamental cluster 
parameters. In Sect.~\ref{struc} we derive structural parameters. In Sect.~\ref{Discus} 
we investigate relations among parameters and with respect to their location in the
Galaxy. Concluding remarks are given in Sect.~\ref{Conclu}.

\section{The sample of overlooked Ruprecht clusters}
\label{RecAdd}

We started by examining the optical images of the objects in order to identify
the best candidates to be further studied with 2MASS photometry (Sect.~\ref{2mass}). 
In short, the selection criteria were {\em (i)} a relatively conspicuous stellar 
density excess (typical of a star cluster), {\em (ii)} a foreground/background 
not excessively dense, and {\em (iii)} an angular size larger than $\sim1\arcmin$ 
(a smaller size might indicate an excessively distant object, too faint to be 
detected with 2MASS). The goal was to focus on objects that would result in reliable 
determinations of fundamental (Sect.~\ref{DFP}) and structural (Sect.~\ref{struc}) 
parameters. The above search resulted in 12 unstudied OCs. Besides these, we also 
included 3 other cases somewhat studied, but with inconsistent parameters 
(Sects.~\ref{DFP} and \ref{struc}).

The sample of Ruprecht clusters that came up from the above search is shown in 
$5\arcmin\times5\arcmin$ B images (Figs.~\ref{fig1}-\ref{fig2}), taken from 
LEDAS\footnote{Leicester Database and Archive Service (LEDAS) DSS/DSS-II service 
on ALBION; {\em http://ledas-www.star.le.ac.uk/DSSimage}.}. Note that the images 
are centred on different coordinates (Table~\ref{tab1}) than those given in WEBDA. 
By default, we start the 2MASS analyses by assuming the WEBDA coordinates as cluster 
centre. However, in most cases the RDPs built after field decontamination - to maximise 
membership probability (Sect.~\ref{struc}), presented a dip in the innermost radial 
bin. So, the central coordinates were computed again to match the absolute maximum 
in the stellar surface density (e.g. Fig.~\ref{fig3}). In most cases, the 
difference between the original and recomputed central coordinates is slight, of 
the order of 1\arcmin, thus well within the estimated optical diameter (col.~7 of
Table~\ref{tab1}).

The images match our selection criteria in different degrees. For instance, the
highest compactness levels occur with Ru\,35, 37, 41, 54, 60, 63, and 152. Ru\,1, 
10, 23, and 27 are projected on the least contaminated fields, with Ru\,174 in the 
densest. Note that the bright (foreground) star 6\,Pup (K\,3\,III) is located at 
$\approx2\arcmin$ to the northwest of the central coordinates of Ru\,37. 

\section{Photometric analysis}
\label{2mass}

Photometry for the sample clusters was extracted from VizieR\footnote{\em
http://vizier.u-strasbg.fr/viz-bin/VizieR?-source=II/246} in a circular field 
of radius $\rx=60\arcmin$. Such a wide extraction area provides the required 
statistics for the determination of the background level (Sect.~\ref{struc}) 
and the colour/magnitude characterisation of the field stars (see below). To 
preserve the photometric quality and, at the same time, work with a statistically 
significant number of stars, only stars with \jj, \hh, and \ks\ errors lower 
than 0.15\,mag were used. Reddening corrections are based on the absorption 
relations $A_J/A_V=0.276$, $A_H/A_V=0.176$, $A_{K_S}/A_V=0.118$, and 
$A_J=2.76\times\ejh$ given by \citet{DSB2002}, with $R_V=3.1$, considering the 
extinction curve of \citet{Cardelli89}. 

Except for Ru\,174 (at $\ell\approx78\degr$, near the border between the $1^{st}$ and
$2^{nd}$ quadrants), the remaining clusters belong to the $3^{rd}$ Galactic quadrant 
(Table~\ref{tab1}), which would make field-star contamination a minor issue (e.g. 
\citealt{ProbFSR}). However, given the relatively poorly-populated nature of the 
present sample (Figs.~\ref{fig1} - \ref{fig2}), it is important to take the field-star 
contamination into account to derive more constrained parameters. In particular, we 
wish to work with CMDs in which cluster evolutionary sequences and field stars are 
disentangled.

\begin{figure}
\resizebox{\hsize}{!}{\includegraphics{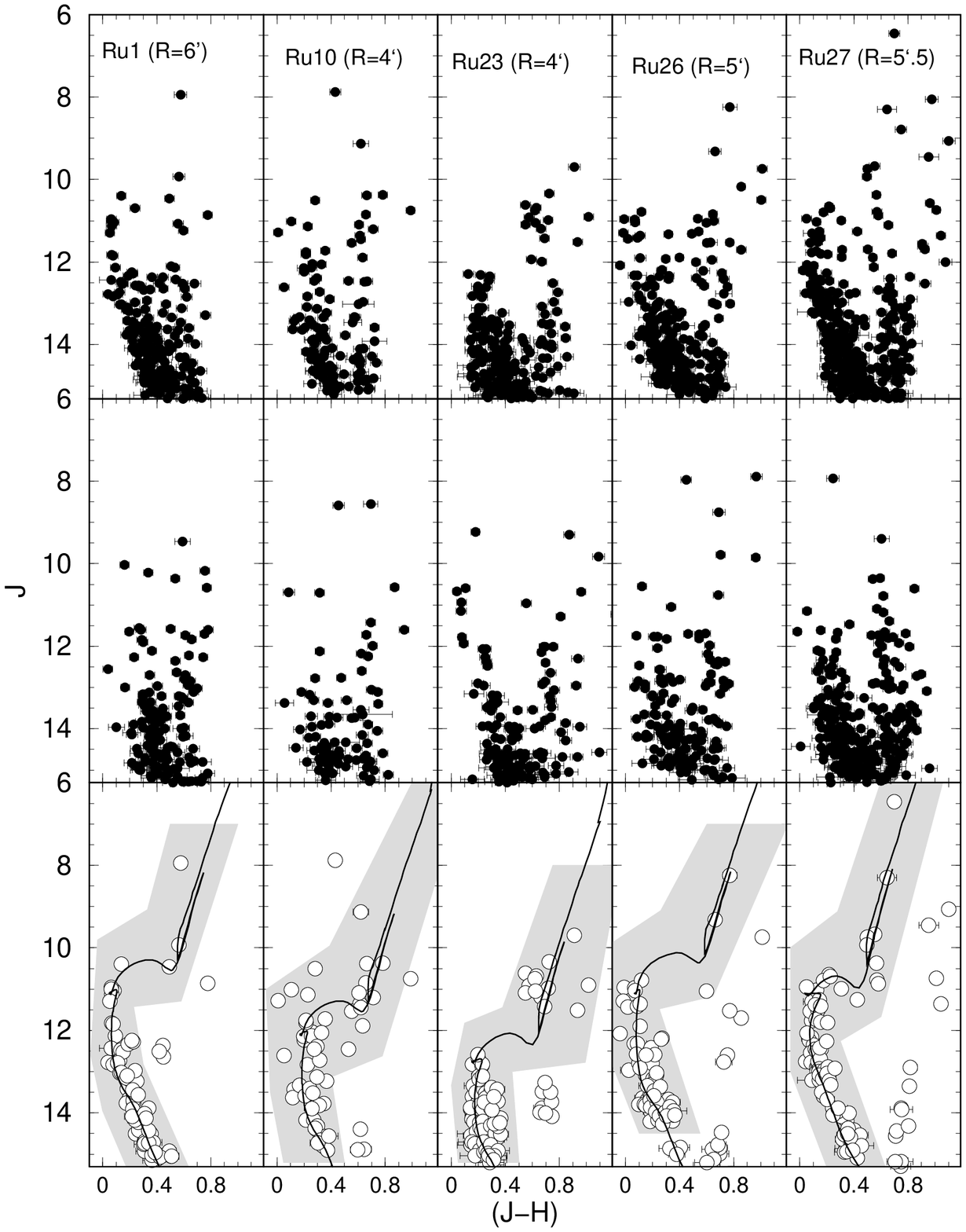}}
\caption[]{$\jj\times\jh$ CMDs of Ru\,1, 10, 23, 26, and 27, showing 
the observed photometry for representative regions (top panels) and the equal-area 
comparison fields (middle). The decontaminated CMDs are shown in the bottom panels, 
together with the isochrone solution (solid line) and colour-magnitude filter (shaded
polygon). Note that, for most stars, the error bars are smaller than the symbol.}
\label{fig3}
\end{figure}

\begin{figure}
\resizebox{\hsize}{!}{\includegraphics{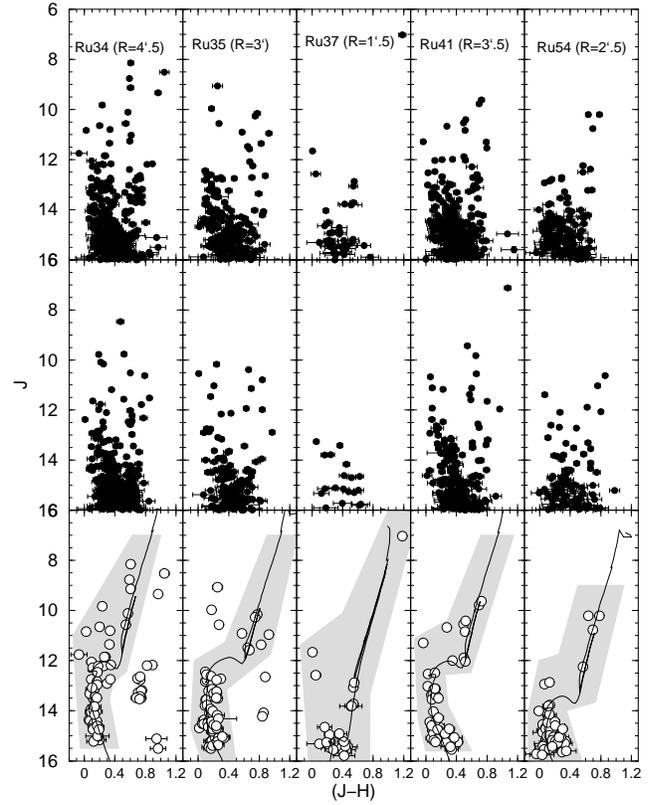}}
\caption[]{Same as Fig.~\ref{fig3} for Ru\,34, 35, 37, 41, and 54.}
\label{fig4}
\end{figure}

\begin{figure}
\resizebox{\hsize}{!}{\includegraphics{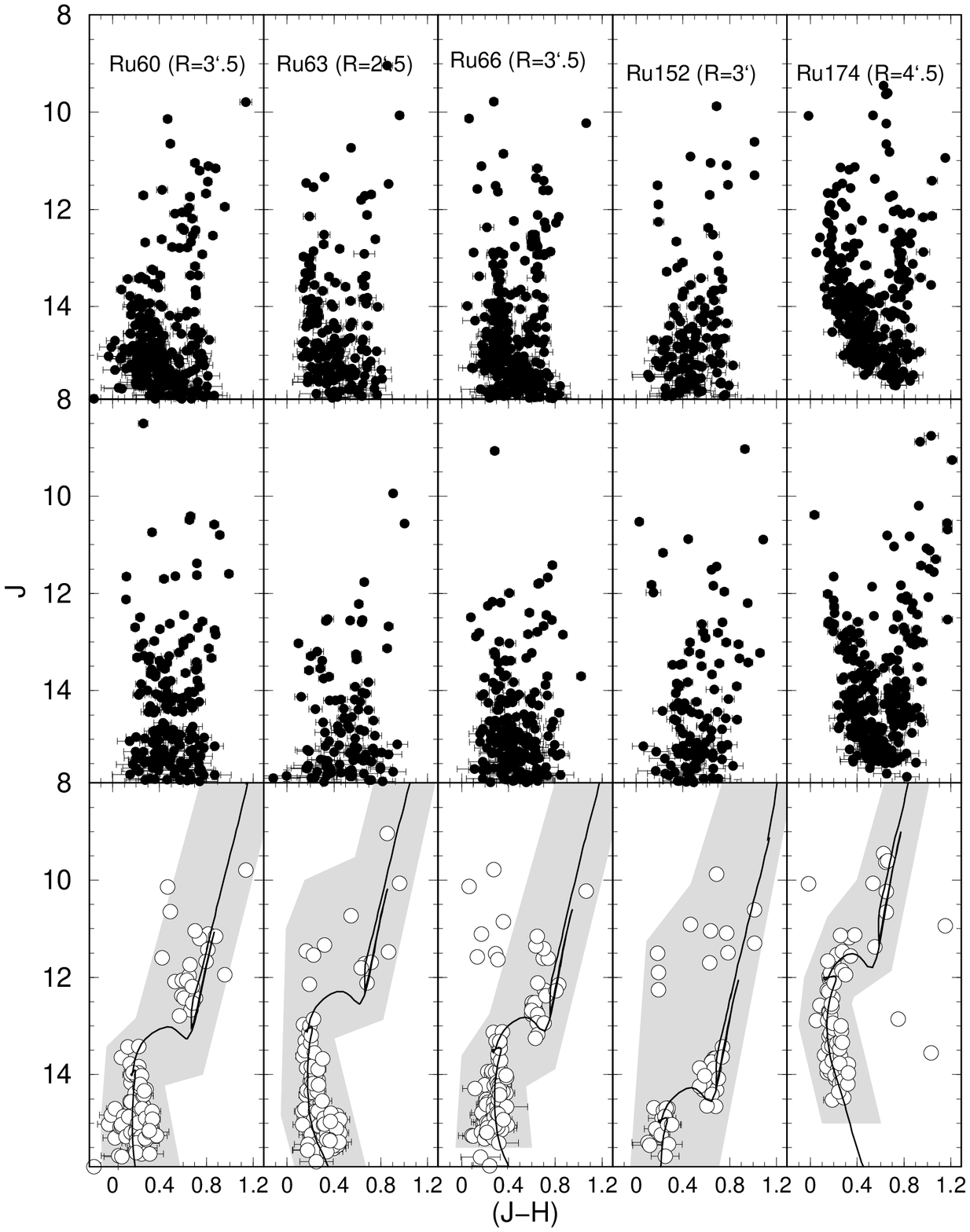}}
\caption[]{Same as Fig.~\ref{fig3} for Ru\,60, 63, 66, 152, and 174.}
\label{fig5}
\end{figure}

For this purpose, we work with a statistical decontamination algorithm that has been 
developed by our group for the proper identification and characterisation of star 
clusters, especially those near the Galactic equator and/or with important fractions 
of faint stars. The algorithm is applied to the 2MASS photometry, which can provide 
the spatial and photometric uniformity required for wide extractions and high star-count 
statistics. 

Working with the wide circular extractions, we start by defining the cluster and 
comparison field regions\footnote{This step is iterative, since we first have to 
build the RDP (Sect.~\ref{struc}) to estimate the cluster size and the location of 
the comparison field. After applying the algorithm, we build the colour-magnitude
filters for the decontaminated cluster CMD. Then we re-build the RDP, re-compute
the cluster size and run again the decontamination algorithm.}. CMDs extracted 
from the cluster region for our objects are shown in Figs.~\ref{fig3}-\ref{fig5}
(top panels). These can be contrasted with the representative (i.e. equal-area)
comparison field CMDs (middle panels). It is important to note that the equal-area 
field extraction is used only for qualitative comparisons, since the algorithm uses 
the wide surrounding area (as defined above) for more statistical representativeness.
Our approach implicitly assumes that the field colour-magnitude distribution {\em (i)} 
is statistically representative of the cluster contaminants, and {\em (ii)} is rather 
spatially uniform. These assumptions are somewhat matched in the $3^{rd}$ Galactic 
quadrant. A detailed description of the decontamination algorithm can be found in 
\citet{BB07} and \citet{vdB92}. For clarity, we provide below only a sketch on how 
it works.

The cluster CMD is divided into a 3D grid of cells with axes along the \jj\ magnitude 
and the \jh\ and \jk\ colours. Then, we compute the probability of a given star to
be found in a particular cell (i.e., for a star with measured magnitude and colour
uncertainties $J\pm\sigma_J$, $\jh\pm\sigma_{(J-H)}$, and $\jk\pm\sigma_{(J-K_S)}$, 
the probability is proportional to the difference between the error function 
computed at the \jj, 
\jh, and \jk-borders of the cell). This step is done for all stars 
and cells, resulting in a number density of $\rm member+field$ stars for each cell 
($\eta_{tot}$). The same steps are applied to the comparison field CMD, from which 
we estimate the field number density ($\eta_{fs}$) for each cell. Next, for each cluster 
cell we subtract the corresponding field number density to obtain a decontaminated 
number density ($\eta_{mem}=\eta_{tot}-\eta_{fs}$). Finally, $\eta_{fs}$ is converted 
back into number of stars and subtracted from each cell, and the $N^{cell}_{clean}$ 
stars that remain in the cell are identified. We also compute the subtraction 
efficiency ($f_{sub}$), which is the sum over all cells of the difference between the 
expected number of field stars (which usually is fractional) and the number of stars 
effectively subtracted (integer). In all cases we obtained $f_{sub}>90\%$.

The above procedure is repeated for 729 different setups (allowing for variations on 
cell size and grid positioning). Each setup produces a total number of member stars 
$N_{mem}=\sum_{cell}N^{cell}_{clean}$, from which we compute the expected total 
number of member stars $\left<N_{mem}\right>$ by averaging out $N_{mem}$ over all 
setups. Stars (identified above) are ranked according to the number of times they 
survive all runs, and only the $\left<N_{mem}\right>$ highest ranked stars are 
considered cluster members and transposed to the respective decontaminated CMD.
The decontaminated $\jj\times\jh$ CMDs of the present sample are shown in 
Figs.~\ref{fig3}-\ref{fig5} (bottom panels).

\begin{figure}
\resizebox{\hsize}{!}{\includegraphics{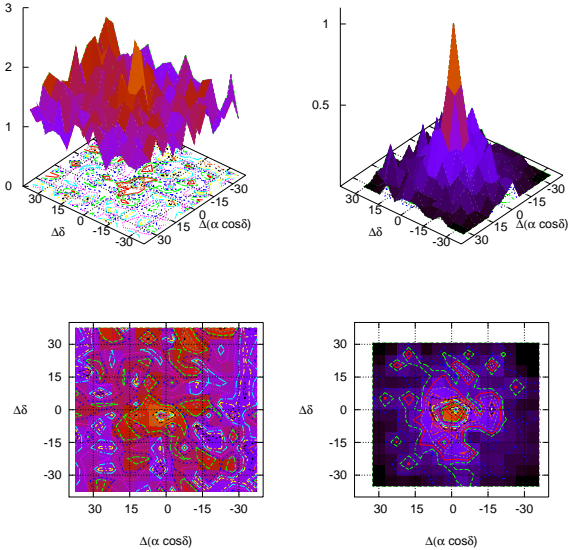}}
\caption[]{2D-perspective on Ru\,1's field decontamination. Top: stellar surface 
density $\sigma(\rm stars\ arcmin^{-2})$ computed before (left panel) and after 
(right) decontamination. Bottom: the respective isopleths. $\Delta(\alpha~\cos(\delta))$ 
and $\Delta\delta$ in arcmin.}
\label{fig6}
\end{figure}

Our decontamination approach relies upon differences in the stellar surface density
measured in CMD cells of separate (cluster and comparison field) spatial regions. 
For a star cluster, which can be characterised by a single-stellar population 
projected 
against a (rather uniform) Galactic stellar field, the decontaminated surface density 
is expected to present a marked excess at the assumed cluster position. We illustrate 
this point by means of the stellar surface density ($\sigma$, in units of 
$\rm stars~arcmin^{-2}$) of Ru\,1 (Fig.~\ref{fig6}), before (top-left panel) and after 
(top-right) decontamination. The respective isopleths are also shown (bottom), in which 
cluster size and geometry can be appreciated. Clearly, the decontamination largely
enhanced the cluster/background contrast, revealing a marked central excess in the
surface density distribution, together with a well-defined, approximately round and 
narrow stellar distribution.

%The surface density is computed in a rectangular mesh with cells 
%$2.5\arcmin\times2.5\arcmin$ wide, reaching total offsets of 
%$|\Delta\alpha~\cos(\delta)|=|\Delta\delta|\approx45\arcmin$ with 
%respect to the cluster centre (Table~\ref{tab1}).

\section{Derivation of fundamental parameters}
\label{DFP}

The presence of somewhat distant and evolved (in different degrees) OCs is suggested 
by the $\jj\times\jh$ CMDs built with the raw photometry of the sample clusters (top 
panels of Figs.~\ref{fig3}-\ref{fig5}). This is clearly confirmed in the decontaminated 
CMDs (bottom panels), in which conspicuous giant clumps and red giant branches can be 
seen in some cases (e.g. Ru\,10, 23, 60, and 66).

We derive the fundamental (reddening, age and distance from the Sun) parameters by
means of the decontaminated CMD morphologies together with Padova isochrones 
(\citealt{Girardi2002}) computed with the 2MASS \jj, \hh, and \ks\ 
filters\footnote{{\em http://stev.oapd.inaf.it/cgi-bin/cmd}. These isochrones are 
very similar to the Johnson-Kron-Cousins ones (e.g. \citealt{BesBret88}), with 
differences of at most 0.01\,mag in colour (\citealt{TheoretIsoc}). }. With respect 
to metallicity, the difference between, e.g. solar and half-solar metallicity isochrones 
for a given age is small, to within the 2MASS photometric uncertainties. Thus, for 
simplicity, we adopt the solar metallicity ones.

Although several analytical approaches for CMD fitting are available (see a summary
in \citealt{NJ06}), we employ a more direct comparison of the isochrones with the 
decontaminated CMD morphology. Specifically, the fits are made {\em by eye}, using 
the combined main sequence (MS) and evolved stellar distributions as constraint. We 
also take variations due to photometric uncertainties into account (which, because of 
the restrictions imposed in Sect.~\ref{2mass}, are usually small) and the presence of 
binaries (which tend to produce a redwards bias in the MS). Starting with the 
isochrones set for zero distance modulus and reddening, we shift them in magnitude 
and colour until a satisfactory match\footnote{Any isochrone solution that occurs 
within the photometric error bars is taken as acceptable.} with the CMD is obtained. 
The {\em best-fits}, 
according to this approach, are shown in Figs.~\ref{fig3}-\ref{fig5} (bottom panels), 
and the respective fundamental parameters are given in Table~\ref{tab1}.

We derive ages within 400\,Myr - 1\,Gyr, except for Ru\,37, which seems to be
somewhat older, with an age of $\sim3$\,Gyr. As expected of optical clusters, the
reddening values are relatively low, $\ebv\le0.9$ (or $\aV\le2.8$). In general,
they are distant from the Sun ($1.5\le\ds(\rm kpc)\le8.0$), and located outside
the solar circle ($0.8\le\Delta R_{SC}(\rm kpc)\le5.0$), except for Ru\,174 at 
$\Delta R_{SC}\sim-0.13$\,kpc.

Our sample has 3 clusters in common with \citet{Kharch05}. Within the uncertainties,
both works agree on the age of Ru\,1 (500 - 600\,Myr); however, we find values of
reddening and distance from the Sun about 60\% higher. For Ru\,26 and Ru\,27
they find ages significantly younger than the present paper, especially for
Ru\,27 with 30\,Myr as compared to 400\,Myr. Reddening and distance from the Sun
also present significant discrepancies (Table~\ref{tab1}). A probable source for
such differences is the presence of unaccounted for field stars in the analysis of 
\citet{Kharch05}. As can be seen in the decontaminated CMDs of Ru\,1, Ru\,26, and 
Ru\,27 (Fig.~\ref{fig3}), the age (and consequently the reddening and distance) is 
quite constrained, to within the quoted errors in Table~\ref {tab1}.

\section{Structural parameters}
\label{struc}

Based on the decontaminated CMD morphologies and corresponding isochrone
solutions (Figs.~\ref{fig3}-\ref{fig5}), we build a colour-magnitude filter
for each cluster. By excluding stars with colours not compatible with those 
of the cluster\footnote{Note that the colour-magnitude filters are wide enough 
to take photometric uncertainties and binaries into account (or other multiple 
systems).}, noise in the RDPs is minimised, while the contrast with the background 
is enhanced (e.g. \citealt{BB07}).

\begin{table*}
\caption[]{Structural parameters derived from the RDPs}
\label{tab2}
%\tiny
\renewcommand{\tabcolsep}{5.0mm}
\renewcommand{\arraystretch}{1.25}
\begin{tabular}{ccccccccc}
\hline\hline
Cluster&$\sigma_0$&\rc&\rl&$1\arcmin$&$\sigma_0$&\rc&\rl\\
       &$\rm(*\,\arcmin^{-2})$&(\arcmin)&(\arcmin) &(pc)&$\rm(*\,pc^{-2})$&(pc)&(pc)\\
(1)&(2)&(3)&(4)&(5)&(6)&(7)&(8)\\
\hline
Ru1  &$16.5\pm8.4$&$0.32\pm0.10$&$8.0\pm1.0$&0.499&$66.1\pm33.8$&$0.16\pm0.05$&$4.0\pm1.0$\\

Ru10 &$13.1\pm6.4$&$0.39\pm0.12$&$4.0\pm0.5$&0.679&$28.4\pm13.9$&$0.26\pm0.08$&$2.7\pm0.3$\\

Ru23 &$8.2\pm2.9$&$0.79\pm0.20$&$6.0\pm0.5$&0.888&$10.3\pm3.7$&$0.70\pm0.18$&$5.3\pm0.4$\\  

Ru26 &$12.5\pm5.4$&$0.42\pm0.13$&$5.5\pm0.5$&0.527&$45.0\pm19.4$&$0.22\pm0.07$&$2.9\pm0.3$\\  

Ru27 &$3.6\pm1.6$&$1.05\pm0.37$&$5.5\pm0.5$&0.433&$19.0\pm8.5$&$0.45\pm0.16$&$2.4\pm0.2$\\ 

Ru34 &$15.6\pm6.4$&$0.34\pm0.08$&$4.5\pm0.5$&0.763&$26.8\pm11.0$&$0.26\pm0.06$&$3.4\pm0.4$\\  

Ru35 &$55.3\pm28.8$&$0.24\pm0.08$&$3.0\pm0.5$&1.135&$42.9\pm22.3$&$0.27\pm0.09$&$3.4\pm0.6$\\  

Ru37 &$34.1\pm15.0$&$0.20\pm0.06$&$1.4\pm0.2$&1.522&$14.7\pm6.5$&$0.30\pm0.09$&$2.1\pm0.3$\\ 

Ru41 &$6.0\pm3.5$&$0.47\pm0.20$&$3.5\pm0.5$&0.913&$7.2\pm4.2$&$0.43\pm0.18$&$3.2\pm0.5$\\  

Ru54 &$45.9\pm29.9$&$0.29\pm0.15$&$3.0\pm0.5$&1.586&$18.3\pm11.9$&$0.46\pm0.24$&$4.7\pm0.8$\\ 

Ru60 &$45.5\pm20.4$&$0.39\pm0.11$&$4.0\pm0.5$&1.786&$14.3\pm6.4$&$0.70\pm0.20$&$7.1\pm0.9$\\ 

Ru63 &$13.1\pm5.2$&$0.90\pm0.25$&$7.0\pm1.0$&1.090&$11.0\pm4.3$&$0.98\pm0.27$&$7.6\pm1.1$\\  

Ru66 &$15.9\pm6.5$&$0.64\pm0.18$&$4.0\pm0.5$&1.091&$6.2\pm2.5$&$0.99\pm0.29$&$6.4\pm0.8$\\

Ru152&$47.0\pm32.0$&$0.20\pm0.08$&$3.5\pm0.5$&2.325&$8.7\pm5.6$&$0.47\pm0.19$&$8.1\pm1.2$\\

Ru174&$5.0\pm1.7$&$0.94\pm0.24$&$4.8\pm0.5$&0.613&$13.4\pm0.5$&$0.58\pm0.15$&$2.9\pm0.3$\\
\hline
\end{tabular}
\begin{list}{Table Notes.}
\item Col.~6: arcmin to parsec scale. For comparison with other clusters, the King-like 
central stellar density ($\sigma_0$) and core radius (\rc), together with the cluster 
radius (\rl), are given both in angular and absolute units. 
\end{list}
\end{table*} 

\begin{figure}
\resizebox{\hsize}{!}{\includegraphics{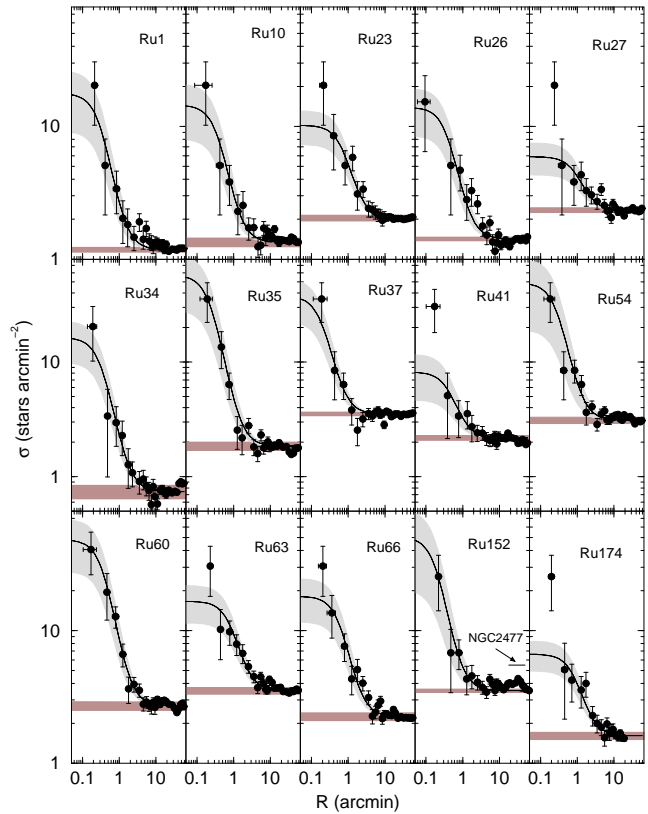}}
\caption[]{Stellar RDPs (filled circles) together with the best-fit King-like profile 
(solid line), the $1\sigma$ uncertainty (light-shaded region) and the residual background 
level (shaded polygon). Note the pronounced central cusps in Ru\,23, 27, 41, 63 and 174. 
The excess in the RDP of Ru\,152 at $R\approx20\arcmin-40\arcmin$ is due to NGC\,2477.}
\label{fig7}
\end{figure}

\begin{table*}
\caption[]{Integrated magnitude and colours}
\label{tab3}
%\tiny
\renewcommand{\tabcolsep}{0.6mm}
\renewcommand{\arraystretch}{1.25}
\begin{tabular}{lccccccccccc}
\hline\hline
       &\multicolumn{4}{c}{Apparent}&\multicolumn{6}{c}{Absolute/reddening corrected}\\
       \cline{2-5}\cline{7-12}\\
Cluster&$m_J$&\jh&\jk&\hk&&$M_J$&$M_V$&$\jh_0$&$\jk_0$&$\hk_0$&$(V-J)_0$\\
(1)&(2)&(3)&(4)&(5)&&(6)&(7)&(8)&(9)&(10)&(11)\\
\hline
Ru\,1  &$8.08\pm0.03$&$+0.30\pm0.06$&$+0.37\pm0.06$&$+0.06\pm0.08$&&$-3.3\pm0.2$&$-1.8\pm0.3$&$+0.22\pm0.06$&$+0.24\pm0.06$&$+0.02\pm0.08$&$1.5\pm0.4$\\
Ru\,10 &$8.67\pm0.02$&$-0.17\pm0.08$&$-0.64\pm0.10$&$-0.47\pm0.13$&&$-3.7\pm0.2$&$-2.2\pm0.4$&$-0.37\pm0.08$&$-0.95\pm0.10$&$-0.58\pm0.13$&$1.6\pm0.4$\\
Ru\,23 &$7.94\pm0.01$&$+0.65\pm0.03$&$+0.84\pm0.02$&$+0.19\pm0.03$&&$-4.9\pm0.2$&$-3.3\pm0.4$&$+0.48\pm0.03$&$+0.58\pm0.02$&$+0.09\pm0.03$&$1.6\pm0.4$\\
Ru\,26 &$9.03\pm0.06$&$-0.21\pm0.26$&$-1.44\pm0.62$&$-1.23\pm0.67$&&$-2.6\pm0.2$&$-1.1\pm0.3$&$-0.32\pm0.26$&$-1.61\pm0.62$&$-1.29\pm0.67$&$1.5\pm0.4$\\
Ru\,27 &$6.42\pm0.03$&$+0.49\pm0.06$&$+0.67\pm0.07$&$+0.17\pm0.08$&&$-4.5\pm0.2$&$-2.9\pm0.4$&$+0.49\pm0.06$&$+0.65\pm0.07$&$+0.17\pm0.08$&$1.6\pm0.4$\\
Ru\,34 &$7.40\pm0.01$&$+0.48\pm0.03$&$+0.63\pm0.02$&$+0.15\pm0.03$&&$-4.7\pm0.2$&$-3.1\pm0.4$&$+0.48\pm0.03$&$+0.63\pm0.02$&$+0.15\pm0.03$&$1.6\pm0.4$\\
Ru\,35 &$9.20\pm0.02$&$+0.49\pm0.03$&$+0.63\pm0.03$&$+0.14\pm0.03$&&$-4.1\pm0.3$&$-2.6\pm0.4$&$+0.35\pm0.03$&$+0.41\pm0.03$&$+0.05\pm0.03$&$1.6\pm0.5$\\
Ru\,37 &$7.04\pm0.02$&$+1.18\pm0.05$&$+1.86\pm0.03$&$+0.68\pm0.05$&&$-6.6\pm0.3$&$-4.9\pm0.5$&$+1.18\pm0.05$&$+1.86\pm0.03$&$+0.68\pm0.05$&$1.7\pm0.5$\\
Ru\,41 &$9.41\pm0.03$&$-0.07\pm0.07$&$-0.51\pm0.10$&$-0.43\pm0.11$&&$-3.2\pm0.3$&$-1.6\pm0.4$&$-0.11\pm0.07$&$-0.57\pm0.10$&$-0.46\pm0.11$&$1.5\pm0.5$\\
Ru\,54 &$8.81\pm0.02$&$+0.60\pm0.02$&$+0.78\pm0.02$&$+0.18\pm0.02$&&$-5.0\pm0.3$&$-3.4\pm0.4$&$+0.56\pm0.02$&$+0.72\pm0.02$&$+0.16\pm0.02$&$1.6\pm0.5$\\
Ru\,60 &$9.29\pm0.03$&$+0.50\pm0.06$&$+0.63\pm0.06$&$+0.14\pm0.08$&&$-5.2\pm0.3$&$-3.6\pm0.4$&$+0.30\pm0.06$&$+0.32\pm0.06$&$+0.02\pm0.08$&$1.6\pm0.5$\\
Ru\,63 &$8.76\pm0.03$&$-0.31\pm0.13$&$-1.42\pm0.35$&$-1.11\pm0.37$&&$-4.6\pm0.2$&$-3.0\pm0.4$&$-0.50\pm0.13$&$-1.72\pm0.35$&$-1.22\pm0.37$&$1.6\pm0.4$\\
Ru\,66 &$9.76\pm0.02$&$+0.05\pm0.05$&$-0.04\pm0.07$&$-0.09\pm0.08$&&$-3.9\pm0.3$&$-2.3\pm0.4$&$-0.23\pm0.05$&$-0.48\pm0.07$&$-0.26\pm0.08$&$1.6\pm0.5$\\
Ru\,152&$9.69\pm0.03$&$+0.41\pm0.06$&$+0.39\pm0.06$&$-0.01\pm0.07$&&$-5.4\pm0.3$&$-3.8\pm0.4$&$+0.20\pm0.06$&$+0.06\pm0.06$&$-0.13\pm0.07$&$1.6\pm0.5$\\
Ru\,174&$8.60\pm0.03$&$-0.56\pm0.08$&$-1.56\pm0.21$&$-1.00\pm0.22$&&$-3.3\pm0.2$&$-1.8\pm0.3$&$-0.66\pm0.08$&$-1.72\pm0.21$&$-1.05\pm0.22$&$1.5\pm0.4$\\
\hline
\end{tabular}
\begin{list}{Table Notes.}
\item Magnitude and colours have been computed with the decontaminated photometry for the
region $R\le\rl$ (Table~\ref{tab2}). Reddening and distance from the Sun (for the absolute
magnitude and reddening-corrected colours - cols.~6-9) are derived in Sect.~\ref{DFP}. 
Cols.~7 and 11: estimated $M_V$ and $(V-J)_0$ (Sect.~\ref{Int_CM}).
\end{list}
\end{table*}

To preserve spatial resolution along the full radial range and, at the same
time keeping moderate error bars, the RDPs are built in rings of increasing 
width with distance from the cluster centre. The set of ring widths used is 
$\Delta\,R=0.25,\ 0.5,\ 1.0,\ 2.5,\ {\rm and}\ 5\arcmin$, respectively for 
$0\arcmin\le R<0.5\arcmin$, $0.5\arcmin\le R<2\arcmin$, $2\arcmin\le R<5\arcmin$, 
$5\arcmin\le R<20\arcmin$, and $R\ge20\arcmin$. Obviously, for any magnitude
range, field stars with the same colour as the cluster's will not be excluded
by the above filtering process. This gives rise to a residual background level,
which can be measured as the average number density of stars away from the
cluster. The $R$ coordinate (and uncertainty) of each ring corresponds to the 
average position (and standard deviation) of the stars inside the ring. The 
resulting RDPs (and residual background) are shown in Fig.~\ref{fig7}. Note
that Ru\,152 is located at $\approx30\arcmin$ to the northwest of NGC\,2477,
which causes a conspicuous bump in the RDP.

We also estimate the cluster radius (\rl) by measuring the distance from the 
centre where the cluster RDP and residual background are statistically 
indistinguishable. In this sense, \rl\ can be considered as an observational 
truncation radius, whose value depends both on the radial distribution of member 
stars and the field density. 

The above RDPs are fitted with the function $\sigma(R)=\sigma_{bg}+\sigma_0/(1+(R/R_c)^2)$, 
where $\sigma_0$ and $\sigma_{bg}$ are the central and residual background stellar 
densities, and \rc\ is the core radius. When applied to star counts, this function 
is similar to that used by \cite{King1962} to the surface-brightness profiles in 
the central parts of globular clusters. Degrees of freedom are minimised by allowing 
only $\sigma_0$ and \rc\ to vary in the fits, while $\sigma_{bg}$ is previously 
measured in the surrounding field and kept fixed. The best-fit solutions are shown 
in Fig.~\ref{fig7}, and the corresponding structural parameters are given in 
Table~\ref{tab2}.

Within uncertainties, the adopted King-like function provides a reasonable description
along the full radial range of the RDPs for most of the sample (Fig.~\ref{fig7}). 
The exceptions are Ru\,27, Ru\,41, and Ru\,174, which present a pronounced cusp (density 
excess over the King-like fit) in the innermost RDP bin. The same appears to apply to 
Ru\,23 and Ru\,63, although only at the $1\sigma$ level. This feature 
has been attributed to a post-core collapse structure in some globular clusters (e.g. 
\citealt{TKD95}). However, such a dynamical evolution-related feature has also been 
detected in the RDP of some Gyr-old OCs, e.g. NGC\,3960 (\citealt{N3960}) and LK\,10 
(\citealt{LKstuff}). Alternatively, clusters that form dynamically cool and with 
significant substructure will probably develop an irregular central region, unless 
such a region collapses and smooths-out the initial substructure (\citealt{Allison09}).

Compared to the distribution of core radii derived for a sample of relatively nearby 
OCs by \citet{Piskunov07} - their Fig.~3, the present clusters occupy the small-\rc\ 
tail. However, we find significant differences, especially in \rc, for the 3 clusters 
in common with \citet{Kharch05}. While they find angular values of \rl\ about twice 
those we derive, their angular values of \rc\ are $\approx5$ (Ru\,27) and $\approx15$ 
(Ru\,1 and Ru\,26) times larger. This, in turn, would imply core radii of the order of 
$\sim2.5$\,pc to $\sim4.0$\,pc, bigger than most of the Galactic globular clusters 
(see, e.g. Fig.~8 of \citealt{Struc11GCs}). These discrepancies probably arise from 
the fact that they do not field-decontaminate their photometry. Because of the 
low-contrast RDPs that result when field stars are not eliminated, structural radii 
derived from RDP fits may not be robust. 

Finally, according to \citet{StrucPar}, the depth-limited 2MASS photometry has only 
a small effect on the core radius determination (by means of the King-like fit), but 
may be somewhat more important for \rl, especially in Ru\,37, for which stars fainter 
than the main-sequence turnoff (MSTO) are not detected.

\section{Discussion} 
\label{Discus}

In the previous sections we derived a set of fundamental and structural parameters 
for a sample of 15 overlooked Ruprecht clusters. Now we use these parameters for
comparison with the previously analysed Ruprecht clusters listed in WEBDA, as well
as to investigate relations among parameters.

\begin{figure}
\resizebox{\hsize}{!}{\includegraphics{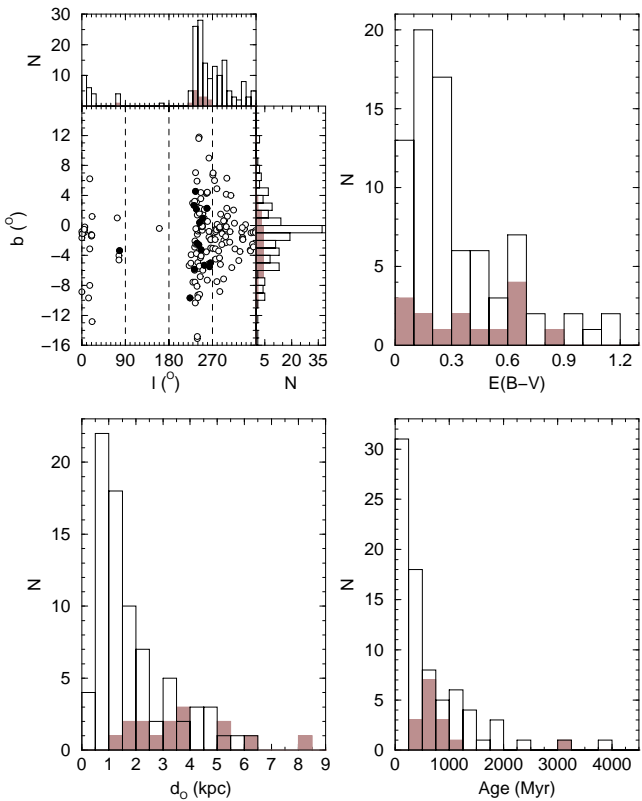}}
\caption{General properties of the present OCs (filled circles and shaded histograms) 
compared to the Ruprecht clusters (empty circles and histograms) listed in WEBDA. 
Dashed lines in the $\ell\times b$ diagram show the borders of the Galactic quadrants.}
\label{fig8}
\end{figure}

\subsection{Comparison with the Ruprecht OCs in WEBDA}
\label{CompRC}

WEBDA contains 171 such clusters with coordinates, but only 79 have age, 
reddening and distance from the Sun been determined so far.

The $\ell$ and $b$ distribution of the Ruprecht clusters in WEBDA is shown
in Fig.~\ref{fig8}. By far, most of them are located in the $3^{rd}$ and
$4^{th}$ quadrants, and within $|b|\la8\degr$. Our sample shares the same
$b$ distribution but is restricted essentially to the $3^{rd}$ quadrant. With
respect to the age and reddening distributions, our sample basically maps those
of the WEBDA clusters. However, our sample is biased towards larger values of
the distance from the Sun, which is consistent with the fact that they still 
haven't been studied in detail.

\subsection{Location in the Galaxy}
\label{LocGal}

\begin{figure}
\resizebox{\hsize}{!}{\includegraphics{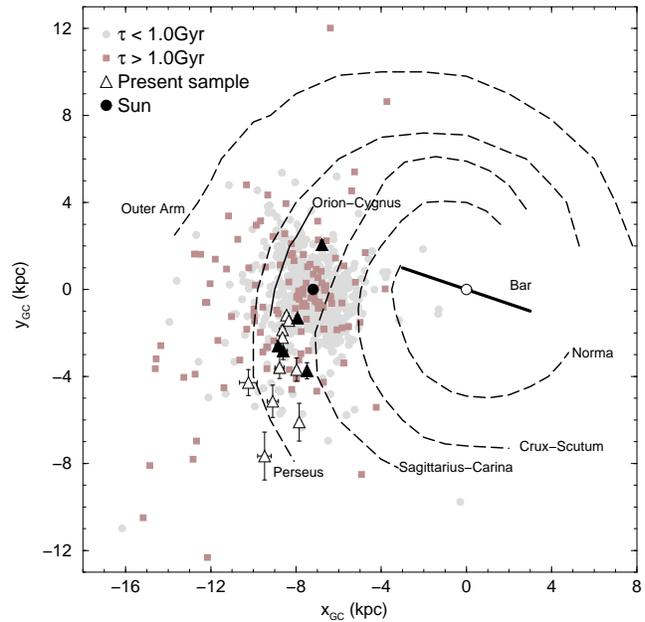}}
\caption{Schematic projection of the Galaxy, as seen from the North pole, with 7.2\,kpc
as the Sun's distance to the Galactic centre, in which the projected distribution of the 
present Ruprecht star clusters (triangles) is compared to the WEBDA OCs younger (circles) 
and older than 1\,Gyr (squares). Clusters with the central cusp (Fig.~\ref{fig7}) are shown 
as filled triangles. Main Galactic structures are identified.}
\label{fig9}
\end{figure}

The positions of the sample clusters, projected onto the Galactic plane, are shown
in Fig.~\ref{fig9}, which contains the spiral arm structure of the Milky Way based on
\citet{GalStr} and \citet{DrimSper01}, derived from HII regions and molecular clouds
(e.g. \citealt{Russeil03}). The Galactic bar is shown with an orientation of 14\degr\
and 6\,kpc of total length (\citealt{Freuden98}; \citealt{Vallee05}). For comparison
we also include the OCs with age and distance from the Sun given in WEBDA, separated
in two age groups, clusters younger or older than 1\,Gyr. 

The main features that emerge from Fig.~\ref{fig9} are summarised as follows. With
respect to the Sun, all 
directions show a depletion in the number of detected OCs for distances farther 
than $\sim2$\,kpc. This occurs because completeness effects (due to crowding 
and high background levels) together with enhanced disruption rates begin to affect 
critically regions more distant than $\sim2$\,kpc, especially towards the bulge 
(e.g. \citealt{DiskProp}). Given the high dissolution rates in the inner Galaxy 
associated with dynamical interactions with the disk, the tidal pull of the bulge, 
and collisions with giant molecular clouds (e.g. \citealt{Friel95}; \citealt{BLG01}; 
\citealt{OldOCs}), old OCs are found preferentially outside the solar circle, a region 
with relatively low tidal stress. On the other hand, the presence of bright stars in 
young OCs allows them to be detected farther than the old ones, especially towards the 
central Galaxy. It should be noted that WEBDA contains essentially optically-selected
OCs, and when surveys in the near-infrared are conducted - like the present one, an 
increasing number of (basically old) OCs at large distances from the Sun are being 
found. Near-infrared searches, in turn, might minimise the OC incompleteness around 
the Sun, but the problem for large distances would still remain (e.g. \citealt{DiskProp}).

Most of the present sample is located between the Perseus and Sagittarius-Carina 
arms, except for Ru\,37 and especially Ru\,152 that are beyond the Perseus arm.
Given their age range ($\rm400\,Myr - 3\,Gyr$), it is quite possible that
some of them may have suffered tidal stress from the arms, by means of encounters
with giant molecular clouds. Collision with such clouds is another potential
dissolution mechanism, especially for low-mass clusters (e.g. \citealt{Wielen71};
\citealt{Wielen91}; \citealt{Gieles06}; \citealt{GAP07}). Such events might have 
accelerated the dynamical evolution and produced changes in the cluster structure. 
However, we note that there is no difference in the projected positions of the 
clusters that display the central cusp in the RDP (Fig.~\ref{fig7}) with respect 
to the King-like ones. In any case, it would be necessary to re-construct their 
orbits through the Galaxy for a deeper analysis on this issue. 

\subsection{Cluster size dependencies}
\label{CSD}

\begin{figure}
\resizebox{\hsize}{!}{\includegraphics{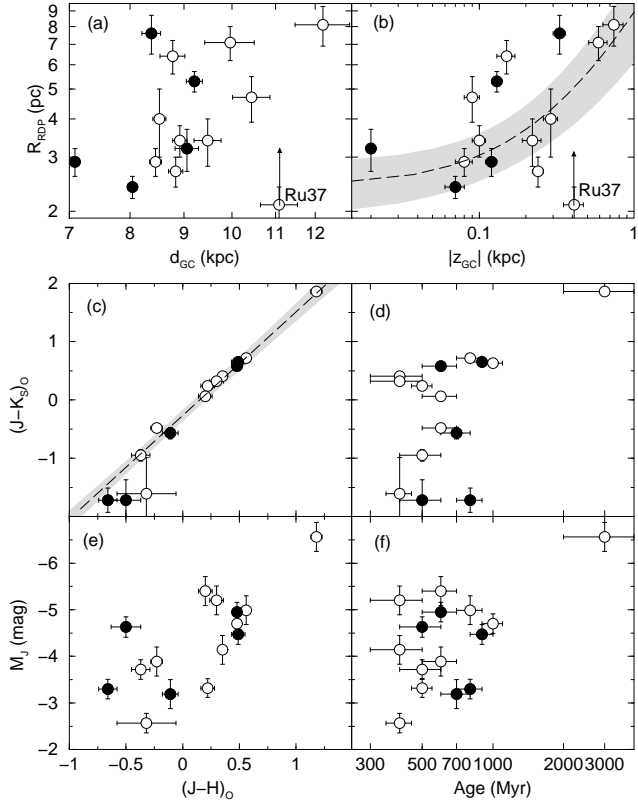}}
\caption{Top: relation of the cluster radius with Galactocentric distance (left)
and distance from the plane (right). Arrows indicate the lower-limit of \rl\ for
Ru\,37. Middle: dependence of the integrated $\jk_0$\ colour on $\jh_0$\ (left) and cluster 
age (right). Bottom: same as above for the absolute \jj\ magnitude. Clusters with 
the central cusp (Fig.~\ref{fig7}) are shown as filled circles. Dashed-line in (b): 
$\rm\rl(pc)=(2.4\pm0.4)+ (6.5\pm2.7)\times|\zgc(kpc)|$. Dashed-line in (c): 
$\jk_0=(-0.27\pm0.04)+ (1.78\pm0.06)\times\jh_0$. Shaded region in (b) and (c): 
$1\sigma$ fit uncertainty.}
\label{fig10}
\end{figure}

Despite the considerable scatter, a first-order dependence of cluster size on 
Galactocentric distance is suggested in Fig.~\ref{fig10} (panel a). Incidentally,
the discordant cluster is Ru\,37 ($\rl\approx2.1$\,pc), the oldest one ($\sim3$\,Gyr) 
of our sample. Given the age and distance from the Sun ($\ds\sim5$\,kpc) of Ru\,37, 
stars fainter than the MSTO are not detected by 2MASS, which may have underestimated
its size. Such a relation has already been observed (e.g. \citealt{Lynga82}; 
\citealt{Tad2002}; \citealt{vdBMP91}), and may reflect the low dissolution rates 
associated with large Galactocentric distances. The rather weak correlation may be
partly due to the fact that our sample clusters are located essentially outside the 
solar circle.

This point can be further investigated by examining the dependence of cluster size
on the vertical distance to the Galactic plane $|\zgc|$, since our clusters are
located within $|\zgc|\la1.0$\,kpc. With the exception of Ru\,37, a somewhat
tight correlation shows up between \rl\ and $|\zgc|$. Indeed, excluding Ru\,37,
the remaining points are roughly described by the relation $\rm\rl(pc)=(2.4\pm0.4)+
(6.5\pm2.7)\times|\zgc(kpc)|$. Again, this relation is consistent with a lower-frequency 
of encounters with giant molecular clouds and the disk for OCs at high $|\zgc|$ with 
respect to those orbiting closer to the plane. However, we note that part of this 
effect may be related to completeness. Given that the average background$+$foreground 
contamination decreases with increasing $|\zgc|$, the external parts of an OC (where 
the surface brightness is intrinsically low) can be detected at larger distances for
high-$|\zgc|$ objects than for those near the plane (\citealt{DiskProp}). On average,
clusters at high-$|\zgc|$ will tend to seem bigger than near the plane.

\subsection{Integrated colours and magnitudes}
\label{Int_CM}

Having decontaminated the photometry (Sect.~\ref{2mass}) and derived structural 
parameters (Sect.~\ref{struc}), we now proceed to compute the integrated (apparent 
and absolute) magnitudes and reddening-corrected colours for the 2MASS bands. Since 
the decontamination efficiency is lower than 100\% (Sect.~\ref{2mass}), we start by
applying the colour-magnitude filter to the decontaminated photometry. Then we sum 
the flux (for a given band) of all stars within $R\le\rl$ (Table~\ref{tab2}) to compute 
the cluster$+$residual field~stars flux ($F_J^{cl+fs}=\sum_i 10^{-0.4J_i}$). The same is
done for all the comparison field stars, to estimate the residual contamination flux 
($F_J^{fs}$). Thus, the integrated magnitude is given by $m_J=-2.5\log\left(F_J^{cl+fs}-
\Omega\times F_J^{fs}\right)$, where $\Omega$ is the ratio between the projected areas 
of the cluster and the comparison field. This procedure is applied to the \jj, \hh, and 
\ks\ bands, and should minimise decontamination efficiency effects.

Since all clusters contain giant and MSTO stars (Figs.~\ref{fig3} - \ref{fig5}) - which 
by far dominate the luminosity, the integrated magnitudes should not be significantly 
affected by the non-detection of the lower-MS stars associated with the depth-limited 
2MASS photometry. Reddening and distance from the Sun (for the absolute magnitude and
reddening-corrected colours) are those computed in Sect.~\ref{DFP}. The results are 
given in Table~\ref{tab3} and discussed below.

The integrated and reddening-corrected $\jk_0$\ and $\jh_0$\ colours are tightly 
related by $\jk_0=(-0.27\pm0.04)+(1.78\pm0.06)\times\jh_0$ (Fig.~\ref{fig10}, panel c), 
and redder clusters also tend to be brighter in \jj\ (e). Panels (d) and (f) seem to 
show a slight tendency of old clusters to be redder and brighter than the young ones. 
This is consistent with most of the stellar luminosity (especially for the more massive
stars) being transferred from the optical to the near-IR, as star clusters become older. 
As a caveat, the latter relation hinges essentially on a single cluster with 
$\sim3$\,Gyr of age. Thus, with the present data we cannot quantify the role of age in 
driving the near-infrared colours (panels c and d), since the age for most of the present 
OCs are restricted to the range $\rm400\,Myr\sim1\,Gyr$.

\begin{figure}
\resizebox{\hsize}{!}{\includegraphics{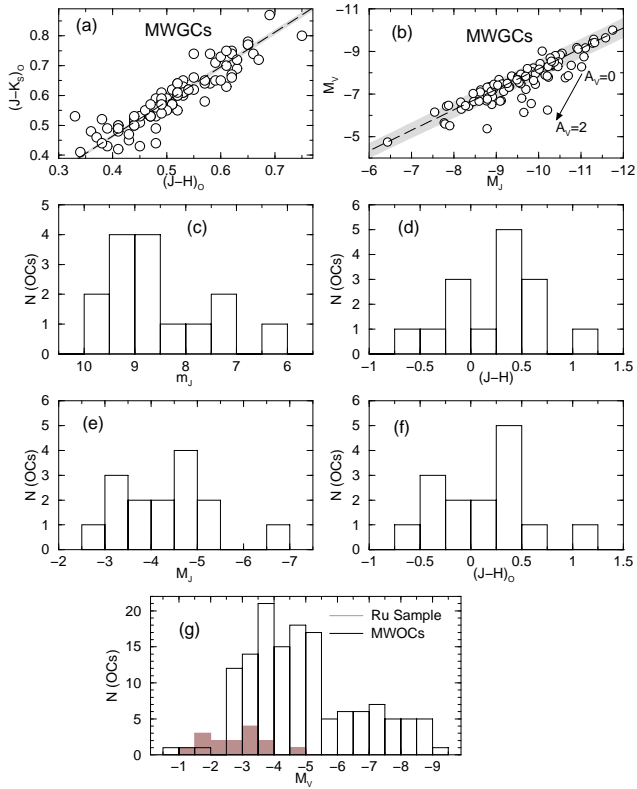}}
\caption{Panels (a) and (b): relations derived from MWGCs for the $\jk_0$\ and $\jh_0$\ 
colours ($\jk_0=(1.16\pm0.01)\times\jh_0$), and the absolute $V$ and $J$ magnitudes
($M_V = (1.41\pm0.27)+ (0.96\pm0.03)\times M_J$). Arrow in (b): reddening vector
for $A_V=2$. Panels (c) and (d): histograms for the number of OCs within bins of 
apparent \jj\ magnitude and \jh\ colour. Panels (e) and (f): same as above for 
the absolute \jj\ magnitude, and the reddening-corrected $\jh_0$\ colour. Panel (g): 
the distribution of the extrapolated $M_V$ values of our Ruprecht sample (shaded 
histogram) compared (empty histogram) to the Galactic OCs of \citet{Lata02} and 
\citet{Bat94}.}
\label{fig11}
\end{figure}

As far as we are aware, there is no study of OC integrated colours or magnitudes 
linking the near-IR to the optical. However, if we assume that OCs and globular 
clusters follow similar scaling relations - at least with respect to integrated 
colours and magnitudes, we can use the Galactic globular clusters (MWGCs) both for 
comparison and to search for such a link. For this purpose, we use the integrated 
and reddening-corrected $(V-\ks)_0$, $\jh_0$, and $\jk_0$ colours computed for a large 
set of MWGCs on 2MASS images (\citealt{Cohen07}), coupled to the respective absolute 
$M_V$ magnitudes from Harris (1996, together with the 2003 revision). The results 
are summarised in Fig.~\ref{fig11} (top panels). Consistent with their old ages, 
the MWGCs $\jk_0$ and $\jh_0$ colours (panel a) are restricted to a narrow range (the 
spread is essentially due to the different metallicities and partly to unaccounted 
for reddening), when compared to the 
set of (younger) Ruprecht OCs (Fig.~\ref{fig10}, panel c). Besides,
the near-IR MWGC colours follow a similar - although somewhat flatter 
($\jk_0\propto1.2\times\jh_0$) - relation than our Ruprecht OCs ($\jk_0\propto1.8\times\jh_0$). 
Finally, with respect to the absolute magnitudes $M_V$ and $M_J$ (panel b), the MWGCs 
are tightly related by $M_V = (1.41\pm0.27)+ (0.96\pm0.03)\times M_J$, over the 
relatively wide ranges $-12\la M_V\la-6.5$ and $-10\la M_J\la-4.5$. Note that the
few discordant points can be accounted for by a reddening under-correction.

The apparent \jj\ magnitude distribution of our sample clusters is clearly biased 
towards fainter objects (panel c of Fig.~\ref{fig11}), with about half of the sample 
having $m_J\ga8.5$, while the absolute \jj\ magnitude (e) is roughly uniformly 
distributed around $\overline{M_J}\approx-4.5$, with a $\pm2.0$\,mag spread. The observed (d) 
and reddening-corrected (f) \jh\ colours are roughly distributed around the average 
value $\overline{\jh}\approx0.5$, with a $\pm1.0$\,mag spread. 

Now, extrapolating the MWGC $M_V\times M_J$ relation to the $M_J$ values derived 
for our Ruprecht clusters, we find $M_V$ values in the range $-5\la M_V\la-1$ 
(panel g of Fig.~\ref{fig11} and Table~\ref{tab3}). Note that the integrated colour
$(V-J)$ is similar in all cases, with an average value $\overline{(V-J)_0}=1.6\pm0.5$. We now
compare the Ruprecht $M_V$ values with those measured for 140 Galactic OCs (MWOCs) 
by \citet{Lata02} together with 106 OCs of \citet{Bat94}\footnote{Both samples have 
similar $M_V$ distributions. For the OCs in common we have used the more recent 
values of \citet{Lata02}.}. Most ($\approx72\%$) of the MWOCs have $M_V$ within 
$-5.5\la M_V\la-2.5$, but the remaining ones can be as luminous as $M_V\approx-10$. 
Clearly, our Ruprecht clusters, in general, appear to be intrinsically faint in the 
optical, with an $M_V$ distribution similar to - but still somewhat biased to lower 
luminosities than -  the MWOCs.

\section{Summary and conclusions}
\label{Conclu}

Given the rather efficient and numerous dissolution mechanisms operating in the 
Galaxy, the majority of the open clusters do not survive beyond a few $10^8$\,yr.
In this context, it is important to investigate the structural and photometric 
properties of OCs that are undergoing this evolved phase.

The present paper focuses on 15 overlooked Ruprecht clusters, 12 of them never before 
studied. With the exception of a single object at $\ell\approx78\degr$,
the remaining clusters are located in the $3^{rd}$ Galactic quadrant, which minimises 
the field-star contamination. We work with 2MASS photometry (with errors $\la0.15$\,mag),
on which we apply field-star decontamination to enhance CMD evolutionary sequences and 
stellar RDPs, thus yielding more constrained fundamental and structural parameters. 

As typical optically discovered clusters, the reddening values are relatively 
low, within $0.0\la\ebv\la0.9$; on the other hand, they are distant from the Sun, 
within $\rm1.5\la\ds(kpc)\la8.0$. The integrated apparent \jj\ magnitudes are rather 
faint, within $6.4\la m_J\la9.8$, but given the distances, the absolute magnitudes
are relatively bright $-6.6\la M_J\la-2.6$. The ages are in the range 400\,Myr - 
1\,Gyr, except for the significantly older Ru\,37, with $\sim3$\,Gyr. The RDPs
are well contrasted with respect to the background and follow the King-like
profile for most of the radial range. Exceptions are Ru\,23, Ru\,27, Ru\,41,
Ru\,63, and Ru\,174, which present a pronounced stellar density excess in the
innermost RDP bin. The core radii of the present sample are small, when compared 
to those of nearby OCs (e.g. \citealt{Piskunov07}). By extrapolating the relation 
between $M_V$ and $M_J$, derived for globular clusters, we estimate $-5\la M_V\la-1$,
which suggests that they are low-luminosity optically-selected clusters.

The sample clusters are located between (or close to) the Perseus and 
Sagittarius-Carina arms, and we present evidence that the cluster size increases 
both with Galactocentric distance and distance to the plane. The latter relation, 
in particular, is consistent with a low frequency of tidal stress associated with
high-$|\zgc|$ regions. 

It is clear from the above analysis that searches for star clusters in catalogues 
of candidates - even in the optical - are far from complete. Detailed investigations 
will certainly add more members to the present-day open cluster census. Besides, since 
young clusters are rather easy to identify even at large distances, the overlooked
clusters are expected to be of the evolved/old age range. Thus, works 
like the present one are important not only because reliable astrophysical parameters 
are derived for a sample of unstudied clusters. Perhaps the main importance lies in the 
unambiguous characterisation of open clusters with ages beyond several $10^8$\,yr. A 
better statistics on the population of these - and older - clusters can be used to 
investigate cluster formation rates and to constrain the time scale of cluster 
dissolution in the Galaxy.

\section*{Acknowledgements}
We thank the reviewer, Dr. A.F. Moffat, for interesting comments and suggestions.
We acknowledge support from the Brazilian Institution CNPq.
This publication makes use of data products from the Two Micron All Sky Survey, which
is a joint project of the University of Massachusetts and the Infrared Processing and
Analysis Centre/California Institute of Technology, funded by the National Aeronautics
and Space Administration and the National Science Foundation. This research has made 
use of the WEBDA database, operated at the Institute for Astronomy of the University
of Vienna.

\label{lastpage}

\begin{thebibliography}{}

\bibitem[\protect\citeauthoryear{Allison et al.}{2009}]{Allison09}
   Allison R.J., Goodwin S.P., Parker R.J., de Grijs R., Portegies Zwart S.F.
   \& Kouwenhoven M.B.N. 2009, ApJL, 700,99
   
\bibitem[\protect\citeauthoryear{Alter et al.}{1970}]{Alter70}
   Alter G., Bal\'azs B., Ruprecht J. \& Vanysek J. 1970, in {\em Catalogue of
   Star Clusters and Associations}, Budapest Akademiai Kiado, 2nd Edition, ed.
   by Alter G., Bal\'azs  B. \& Ruprecht J.
   
%\bibitem[\protect\citeauthoryear{Bastian et al.}{2005}]{Bast05}
%   Bastian N., Gieles M., Lamers H.J.G.L.M., Scheepmaker R.A. \& de Grijs R. 
%   2005, A\&A, 431, 905
   
%\bibitem[\protect\citeauthoryear{Battinelli \& Capuzzo-Dolcetta}{1991}]{Bat91}
%   Battinelli P. \& Capuzzo-Dolcetta R. 1991, MNRAS, 249, 76
   
\bibitem[\protect\citeauthoryear{Battinelli, Brandimarti \& Capuzzo-Dolcetta}{1994}]{Bat94}
   Battinelli P., Brandimarti A. \& Capuzzo-Dolcetta R. 1994, A\&AS, 104, 379
   
\bibitem[\protect\citeauthoryear{Baumgardt \& Makino}{2003}]{BM03}
   Baumgardt H. \& Makino J. 2003, MNRAS, 340, 227
  
%\bibitem[\protect\citeauthoryear{van den Bergh}{1966}]{vdB66}
%   van den Bergh S. 1966, AJ, 71, 990
   
\bibitem[\protect\citeauthoryear{van den Bergh, Morbey \& Pazder}{1991}]{vdBMP91}
   van den Bergh S., Morbey C. \& Pazder J. 1991, ApJ, 375, 594
   
\bibitem[\protect\citeauthoryear{Bergond, Leon \& Guilbert}{2001}]{BLG01}
   Bergond G., Leon S. \& Guilbert J. 2001, A\&A, 377, 462
     
\bibitem[\protect\citeauthoryear{Bessel \& Brett}{1988}]{BesBret88}
   Bessel M.S. \& Brett J.M. 1988, PASP, 100, 1134
   
%\bibitem[\protect\citeauthoryear{Bica, Bonatto \& Dutra}{2008}]{Bochum1}
%   Bica E., Bonatto C. \& Dutra C. 2008, A\&A, 489, 1129

%\bibitem[\protect\citeauthoryear{Bica et al.}{2006}]{GCProp}
%   Bica E., Bonatto C., Barbuy B. \& Ortolani S. 2006, A\&A, 450, 105

\bibitem[\protect\citeauthoryear{Bica, Bonatto \& Camargo}{2008}]{ProbFSR}
   Bica E., Bonatto C. \& Camargo D. 2008, MNRAS, 385, 349

\bibitem[\protect\citeauthoryear{Bonatto, Bica \& Girardi}{2004}]{TheoretIsoc}
   Bonatto C., Bica E. \& Girardi L. 2004, A\&A, 415, 571

\bibitem[\protect\citeauthoryear{Bonatto et al.}{2006}]{DiskProp}
   Bonatto C., Kerber L.O., Bica E. \& Santiago B.X. 2006, A\&A, 446, 121
   
\bibitem[\protect\citeauthoryear{Bonatto \& Bica}{2006}]{N3960}
   Bonatto C. \& Bica E. 2006, A\&A, 455, 931

%\bibitem[\protect\citeauthoryear{Bonatto \& Bica}{2005}]{DetAnalOCs}
%   Bonatto C. \&  Bica E. 2005, A\&A, 437, 483
   
\bibitem[\protect\citeauthoryear{Bonatto \& Bica}{2007a}]{OldOCs}
   Bonatto C. \& Bica E. 2007a, A\&A, 473, 445

%\bibitem[\protect\citeauthoryear{Bonatto, Santos Jr. \& Bica}{2006}]{N6611}
%   Bonatto C., Santos Jr. J.F.C. \& Bica E. 2006, A\&A, 445, 567
   
%\bibitem[\protect\citeauthoryear{Bonatto et al.}{2006}]{N4755}
%   Bonatto C., Bica E., Ortolani S. \& Barbuy B. 2006, A\&A, 453, 121
   

\bibitem[\protect\citeauthoryear{Bonatto \& Bica}{2007b}]{BB07}
   Bonatto C. \& Bica E. 2007b, MNRAS, 377, 1301
   
\bibitem[\protect\citeauthoryear{Bonatto \& Bica}{2008a}]{StrucPar}
   Bonatto C. \& Bica E. 2008a, A\&A, 477, 829
   
\bibitem[\protect\citeauthoryear{Bonatto \& Bica}{2008b}]{Struc11GCs}
   Bonatto C. \& Bica E. 2008b, A\&A, 479, 741
   
\bibitem[\protect\citeauthoryear{Bonatto \& Bica}{2009a}]{LKstuff}
   Bonatto C. \& Bica E. 2009a, MNRAS, 392, 483
   
\bibitem[\protect\citeauthoryear{Bonatto \& Bica}{2009b}]{N2244}
   Bonatto C. \& Bica E. 2009b, MNRAS, 394, 2127
   
%\bibitem[\protect\citeauthoryear{Bonatto \& Bica}{2009c}]{Pi5}
%   Bonatto C. \& Bica E. 2009c, MNRAS, 397, 1915
   
\bibitem[\protect\citeauthoryear{Bonatto \& Bica}{2010}]{vdB92}
   Bonatto C. \& Bica E. 2010, A\&A, in press

\bibitem[\protect\citeauthoryear{Cardelli, Clayton \& Mathis}{1989}]{Cardelli89}
   Cardelli J.A., Clayton G.C. \& Mathis, J.S. 1989, ApJ, 345, 245
   
\bibitem[\protect\citeauthoryear{Carraro, Janes \& Eastman}{2005}]{Carr05}
   Carraro G., Janes K.A. \& Eastman J.D. 2005, MNRAS, 364, 179
   
\bibitem[\protect\citeauthoryear{Carraro et al.}{2006}]{Carr06a}
   Carraro G., Janes K.A., Costa E. \& M\'endez R.A. 2006, MNRAS, 368, 1078
   
\bibitem[\protect\citeauthoryear{Carraro, Subramaniam \& Janes}{2006}]{Carr06b}
   Carraro G., Subramaniam A. \& Janes K.A. 2006, MNRAS, 371, 1301
   
\bibitem[\protect\citeauthoryear{Cohen et al.}{2007}]{Cohen07}
   Cohen J.G., Hsieh S., Metchev S., Djorgovski S.G. \& Malkan M. 
   2007, AJ, 133, 99
   
%\bibitem[\protect\citeauthoryear{Chen, de Grijs \& Zhao}{2004}]{CGZ07}
%   Chen L., de Grijs R. \& Zhao J.L. 2007, AJ, 134, 1368
   
%\bibitem[\protect\citeauthoryear{Clari\'a}{1974a}]{Claria74a}
%   Clari\'a J.J. 1974a, ApJ, 79, 1022

%\bibitem[\protect\citeauthoryear{Clari\'a}{1974b}]{Claria74b}
%   Clari\'a J.J. 1974b, A\&A, 37, 229
   
%\bibitem[\protect\citeauthoryear{Collinder}{1931}]{Coll31}
%   Collinder P. 1931, AnLun, 2, 1

\bibitem[\protect\citeauthoryear{Dias et al.}{2002}]{diasCat}
   Dias W.S., Alessi B.S., Moitinho A. \& L\'epine J.R.D. 2002, A\&A, 389, 871
   
\bibitem[\protect\citeauthoryear{Drimmel \& Spergel}{2001}]{DrimSper01}
   Drimmel R., \& Spergel D.N. 2001, ApJ, 556, 181

\bibitem[\protect\citeauthoryear{Dutra, Santiago \& Bica}{2002}]{DSB2002}
   Dutra C.M., Santiago B.X. \& Bica E. 2002, A\&A, 383, 219
   
\bibitem[\protect\citeauthoryear{Freudenreich}{1998}]{Freuden98}
   Freudenreich H.T. 1998, ApJ, 492, 495
      
\bibitem[\protect\citeauthoryear{Friel}{1995}]{Friel95}
   Friel E.D. 1995, ARA\&A 1995, 33, 381
   
%\bibitem[\protect\citeauthoryear{Froebrich, Scholz \& Raftery}{2007}]{Froeb07}
%   Froebrich D., Scholz A. \& Raftery C.L. 2007, MNRAS, 374, 399
   
%\bibitem[\protect\citeauthoryear{Furlan et al.}{2009}]{Furlan09}
%   Furlan E., Watson D.M., McClure M.K., Manoj, P., Espaillat C., D'Alessio P.,
%   Calvet N., Kim K.H. et al. 2009, ApJ, 703, 1964
   
%\bibitem[\protect\citeauthoryear{Ghez et al.}{2008}]{Ghez08}
%   Ghez A.M., Salim S., Weinberg N.N., Lu J.R., Do T., Dunn J.K., Matthews K.,
%   Morris M.R. et al. 2008, ApJ, 689, 1044
   
\bibitem[\protect\citeauthoryear{Gieles, Athanassoula \& Portegies Zwart}{2007}]{GAP07}
   Gieles M., Athanassoula E. \& Portegies Zwart S.F. 2007, MNRAS, 376, 809

\bibitem[\protect\citeauthoryear{Gieles et al.}{2006}]{Gieles06}
   Gieles M., Portegies Zwart S.F., Baumgardt H., Athanassoula E., Lamers H.J.G.L.M.
   Sipior M. \& Leenaarts J. 2006, MNRAS, 371, 793
   
%\bibitem[\protect\citeauthoryear{Gieles, Sana \& Portegies Zwart}{2008}]{GSPZ10}
%   Gieles M., Sana H. \& Portegies Zwart S.F. 2010, MNRAS, 402, 1750
   
\bibitem[\protect\citeauthoryear{Girardi et al.}{2002}]{Girardi2002}
   Girardi L., Bertelli G., Bressan A., Chiosi C., Groenewegen M.A.T.,
   Marigo P., Salasnich B. \& Weiss A. 2002, A\&A, 391, 195  
     
%\bibitem[\protect\citeauthoryear{Goodwin}{2009}]{GoodW09}
%   Goodwin S.P. 2009, Ap\&SS, 324, 259
   
\bibitem[\protect\citeauthoryear{Goodwin \& Bastian}{2006}]{GoBa06}
   Goodwin S.P. \& Bastian N. 2006, MNRAS, 373, 752
   
\bibitem[\protect\citeauthoryear{Harris}{1996}]{H96}
   Harris W.E. 1996, AJ, 112, 1487
   
%\bibitem[\protect\citeauthoryear{Gouliermis et al.}{2000}]{Goul00}
%   Gouliermis D., Kontizas M., Korakitis R., Morgan D.H., Kontizas E.
%   \& Dapergolas A. 2000, AJ, 119, 1737
   
%\bibitem[\protect\citeauthoryear{de Grijs, Kouwenhoven \& Goodwin}{2008}]{GKG08}
%   de Grijs R., Kouwenhoven M.B.N. \& Goodwin S.P. 2008, AN, 329, 972
   
%\bibitem[\protect\citeauthoryear{de Grijs \& Goodwin}{2008}]{deGG08}
%   de Grijs R. \& Goodwin S.P. 2008, MNRAS, 383, 1000
   
%\bibitem[\protect\citeauthoryear{de Grijs \& Goodwin}{2009}]{deGG09}
%   de Grijs R. \& Goodwin S.P. 2009, in {\em IAU Symposium 256}, van 
%   Loon J.T. \& Oliveira J.M., eds., Cambridge Univ. Press
   
%\bibitem[\protect\citeauthoryear{Gum}{1955}]{Gum55}
%   Gum C.S. 1955, MmRAS, 67, 155
   
\bibitem[\protect\citeauthoryear{Khalisi, Amaro-Seoane \& Spurzem}{2007}]{Khalisi07}
   Khalisi E., Amaro-Seoane P. \& Spurzem R. 2007, MNRAS, 374, 703
   
\bibitem[\protect\citeauthoryear{Kharchenko et al.}{2005}]{Kharch05}        
   Kharchenko N.V., Piskunov A.E., R\"oser S., Schilbach E. \& 
   Scholz R.-D. 2005, A\&A, 438, 1163
   
\bibitem[\protect\citeauthoryear{King}{1962}]{King1962}
   King I. 1962, AJ, 67, 471
   
%\bibitem[\protect\citeauthoryear{Kroupa}{2001}]{Kroupa2001}
%   Kroupa P. 2001, MNRAS, 322, 231
      
\bibitem[\protect\citeauthoryear{Lada \& Lada}{2003}]{LL2003}
   Lada C.J. \& Lada E.A. 2003, ARA\&A, 41, 57
   
\bibitem[\protect\citeauthoryear{Lamers \& Gieles}{2006}]{LG06}
   Lamers H.J.G.L.M. \& Gieles M. 2006, A\&AL, 455, 17
   
\bibitem[\protect\citeauthoryear{Lamers et al.}{2005}]{Lamers05}
   Lamers H.J.G.L.M., Gieles M., Bastian N., Baumgardt H.,
   Kharchenko N.V. \& Portegies Zwart S. 2005, A\&A, 441, 117
   
\bibitem[\protect\citeauthoryear{Lata et al.}{2002}]{Lata02}
   Lata S., Pandey A.K., Sagar R. \& Mohan V. 2002, A\&A, 388, 158
   
%\bibitem[\protect\citeauthoryear{Lauberts}{1982}]{Lauberts82}
%   Lauberts A. 1982, ESO/Uppsala survey of the ESO(B) atlas, European Southern
%   Observatory
   
\bibitem[\protect\citeauthoryear{Lyng\aa}{1982}]{Lynga82}
   Lyng\aa\ G. 1982, A\&A, 109, 213
   
%\bibitem[\protect\citeauthoryear{Magakian}{2003}]{Magakian03}
%   Magakian T.Y. 2003, A\&A, 399, 141
   
%\bibitem[\protect\citeauthoryear{Massey, Johnson \& Gioia-Eastwood}{1995}]{Massey95}
%   Massey P., Johnson K.E. \& De Gioia-Eastwood K. 1995, ApJ, 454, 151

\bibitem[\protect\citeauthoryear{Momany et al.}{2006}]{GalStr}
   Momany Y., Zaggia S., Gilmore G., Piotto G., Carraro G., Bedin
   L.R. \& de Angeli F. 2006, A\&A, 451, 515
   
\bibitem[\protect\citeauthoryear{Naylor \& Jeffries}{2006}]{NJ06}
   Naylor T. \& Jeffries R.D. 2006, MNRAS, 373, 1251
   
\bibitem[\protect\citeauthoryear{Oort}{1958}]{Oort58}
   Oort J.H. 1958, in {\em Ricerche Astronomiche}, 5, 415, Specola Vaticana,
   Proc. of a Conference at Vatican Observatory, Castel Gandolfo, May 20-28, 1957,
   ed. D.J.K. O'Connell
   
\bibitem[\protect\citeauthoryear{Pavani \& Bica}{2007}]{PB07}
   Pavani D.N. \& Bica E. 2007, MNRAS, 468, 139
   
%\bibitem[\protect\citeauthoryear{Pettersson \& Reipurth}{1994}]{PetBo94}
%   Pettersson B. \& Reipurth B. 1994, A\&ASS, 104, 233
   
\bibitem[\protect\citeauthoryear{Piskunov et al.}{2007}]{Piskunov07}
   Piskunov A.E., Schilbach E., Kharchenko N.V., R\"oser S. \& Scholz R.-D.
   2007, A\&A, 468, 151
   
%\bibitem[\protect\citeauthoryear{Poetzel, Mundt \& Ray}{1989}]{Poetzel89}
%   Poetzel R., Mundt R. \& Ray T.P. 1989, A\&A, 224, L13
   
%\bibitem[\protect\citeauthoryear{Rodgers, Campbell \& Whiteoak}{1960}]{RCW60}
%   Rodgers A.W., Campbell C.T. \& Whiteoak, J.B. 1960, MNRAS, 121, 103
   
\bibitem[\protect\citeauthoryear{Russeil}{2003}]{Russeil03}
   Russeil D. 2003, A\&A, 397, 133
   
%\bibitem[\protect\citeauthoryear{Siess, Dufour \& Forestini}{2000}]{Siess2000}
%   Siess L., Dufour E. \& Forestini M. 2000, A\&A, 358, 593

\bibitem[\protect\citeauthoryear{Skrutskie et al.}{1997}]{2mass1997}
   Skrutskie M., Schneider S.E., Stiening R., Strom S.E., Weinberg M.D.,
   Beichman C., Chester T., Cutri R .et al. 1997, in {\it The Impact
   of Large Scale Near-IR Sky Surveys}, ed. F. Garzon et al., Kluwer 
   (Netherlands), 210, 187
   
%\bibitem[\protect\citeauthoryear{Soares \& Bica}{2003}]{Soares03}
%   Soares J.B. \& Bica E. 2003, A\&A, 404, 217
   
\bibitem[\protect\citeauthoryear{Spitzer}{1958}]{Spitzer58}
   Spitzer L. 1958, ApJ 127, 17
   
%\bibitem[\protect\citeauthoryear{Spitzer}{1987}]{Spitzer87}
%   Spitzer L. 1987, in {\em Dynamical Evolution of Globular Clusters},
%   Princeton, NJ, Princeton University Press, p. 191
   
\bibitem[\protect\citeauthoryear{Tadross et al.}{2002}]{Tad2002}
   Tadross A.L., Werner P., Osman A. \& Marie M. 2002, NewAst, 7, 553   
   
\bibitem[\protect\citeauthoryear{Trager, King \& Djorgovski}{1995}]{TKD95}
   Trager S.C., King I.R. \& Djorgovski S. 1995, AJ, 109, 218
   
%\bibitem[\protect\citeauthoryear{Trippe et al.}{2008}]{Trippe08}
%   Trippe S., Gillessen S., Gerhard O.E., Bartko H., Fritz T.K., Maness H.L.
%   Eisenhauer F., Martins F., et al.  2008, A\&A, 492, 419
   
%\bibitem[\protect\citeauthoryear{Tutukov}{1978}]{tutu78}
%   Tutukov A.V. 1978, A\&A, 70, 57
   
\bibitem[\protect\citeauthoryear{Vall\'ee}{2005}]{Vallee05}
   Vall\'ee J.P. 2005, AJ, 130, 56
   
%\bibitem[\protect\citeauthoryear{Vogt \& Moffat}{1973}]{VM73}
%   Vogt N. \& Moffat A.F.J. 1973, A\&AS, 9, 97
   
%\bibitem[\protect\citeauthoryear{}{2007}]{Whit07}
%   Whitmore B.C., Chandar R. \& Fall S.M. 2007, AJ, 133, 1067
   
\bibitem[\protect\citeauthoryear{Wielen}{1971}]{Wielen71}
   Wielen R. 1971, A\&A, 13, 309

\bibitem[\protect\citeauthoryear{Wielen}{1991}]{Wielen91}
   Wielen R. 1991, in ``{\it The Formation and Evolution of Star Clusters}'',
   K. Janes (ed.), Astron. Soc. Pac. Conf. Ser. 13, San Francisco, CA, p. 343-349
   
%\bibitem[\protect\citeauthoryear{Zacharias et al.}{2009}]{Zach09}
%   Zacharias N., Finch C., Girard T., Hambly N., Wycoff G.,
%   Zacharias M.I. , Castillo D., Corbin T. et al. 2009, AJ, submitted
   
\end{thebibliography}
\end{document}